\documentclass[aps,prd,superscriptaddress,twoside,twocolumn,nofootinbib,10pt,%
showpacs,floatfix]{revtex4-1}

\usepackage{amsmath,amssymb}
\usepackage{graphicx,bm}
\usepackage{slashed}
\usepackage{dcolumn}
\usepackage{epstopdf}
\usepackage{ulem} 
\usepackage[usenames]{color}

\allowdisplaybreaks

\begin{document}

\title{Skyrmions with vector mesons in the hidden local symmetry approach}

\author{Yong-Liang Ma}
\email{ylma@hken.phys.nagoya-u.ac.jp}
\affiliation{Department of Physics, Nagoya University,  Nagoya,
464-8602, Japan}

\author{Ghil-Seok Yang}
\email{ghsyang@gmail.com}
\affiliation{Department of Physics, Kyungpook National University,
Daegu 702-701, Korea}

\author{Yongseok Oh}
\email{yohphy@knu.ac.kr}
\affiliation{Department of Physics, Kyungpook National University,
Daegu 702-701, Korea}

\author{Masayasu Harada}
\email{harada@hken.phys.nagoya-u.ac.jp}
\affiliation{Department of Physics, Nagoya University,  Nagoya,
464-8602, Japan}

\date{\today}
\begin{abstract}

The roles of light $\rho$ and $\omega$ vector mesons in the Skyrmion are
investigated in a chiral Lagrangian derived from the hidden local symmetry
(HLS) up to $O(p^4)$ including the homogeneous Wess-Zumino (hWZ) terms.
We write a general ``master formula" that allows us to determine the 
parameters of the HLS Lagrangian from a class of holographic QCD models valid 
at large $N_c$ and $\lambda$ ('t Hooft constant) limit by integrating out the 
infinite towers of vector and axial-vector mesons other than the lowest $\rho$ 
and $\omega$ mesons. 
Within this approach we find that the physical properties of the Skyrmion as 
the solitonic description of baryons are \textit{independent\/} of the HLS 
parameter $a$. 
Therefore the only parameters of the model are the pion decay constant and 
the vector meson mass. 
Once determined in the meson sector, we have a totally parameter-free theory 
that allows us to study unequivocally the role of light vector mesons in the 
Skyrmion structure.
We find, as suggested by Sutcliffe, that inclusion of the $\rho$ meson reduces 
the soliton mass, which makes the Skyrmion come closer to the 
Bogomol'nyi-Prasad-Sommerfield (BPS) soliton, but the role of the $\omega$ 
meson is found to increase the soliton mass.
In a stark contrast, the $\Delta$-$N$ mass difference, which is determined by 
the moment of inertia in adiabatic collective quantization of the Skyrmion, is 
increased by the $\rho$ vector meson, while it is reduced by the inclusion of 
the $\omega$ meson.
All these observations show the importance of the $\omega$ meson in the 
properties of the nucleon and nuclear matter in the Skyrme model.
\end{abstract}


\maketitle


\section{Introduction}
\label{sec:intro}

In accessing dense baryonic matter, one possible approach that unifies both 
the elementary baryons and multi-baryons system was proposed in 
Refs.~\cite{BR,LPMRV03}.
In this approach, starting with a chiral Lagrangian, the single baryon is 
generated as a Skyrmion and multi-Skyrmions are put on crystal lattice to 
simulate many-baryon systems and dense matter.

However, the previous works in this approach suffer from the modeling of the 
effective chiral Lagrangian and determination of the low energy constants 
(LECs).
This problem becomes serious when one considers higher order chiral Lagrangian 
or introduce more mesonic degrees of freedom, which prevents systematic studies
for Skyrmion properties.
On the other hand, when one starts with holographic QCD (hQCD) models and 
integrates out infinite towers of mesons such as vector and axial-vector 
mesons except a few low-lying mesons as done in Refs.~\cite{HMY06,HMY10}, it 
leads to a chiral Lagrangian with the values of all the LECs fixed by only a 
few phenomenological inputs.
Therefore, the number of parameters is drastically reduced, which allows the 
studies on Skyrmions in a systematic way.
In the present article, we use the general master formula derived in this way 
to determine the LECs of the $O(p^4)$ terms of the hidden local symmetry (HLS) 
Lagrangian, which leaves undetermined by theory the pion decay constant, 
vector meson mass, and the HLS parameter $a$. 
As will be shown, all physical observables that we will consider are 
independent of $a$. 
So by fixing the pion decay constant and the vector meson mass from the meson 
sector, we are able to study the Skyrmion properties in a totally 
parameter-independent way, a feat that as far as we know, has not been 
achieved before.
In addition, we work with the chiral Lagrangian including the homogeneous 
Wess-Zumino (hWZ) terms for studying the role of the $\omega$ vector meson in 
the properties of a single Skyrmion.
This is our first step towards more complete studies on Skyrmions for nucleon 
structure and baryonic matter.
The main results of this work were quoted in Ref.~\cite{MOYHLPR12}, and here 
we provide the details of the calculations in this model as well as more 
detailed analyses on the Skyrmion mass and size.

The Skyrme model~\cite{Skyrme61,Skyrme62} is the nonlinear sigma model 
stabilized by the Skyrme term, a four-derivative term, where the baryons 
emerge as stable field configurations with a non-trivial geometrical structure.
It is well accepted that the nonlinear sigma model captures the physics of QCD 
at very low energy scales and as the energy scales increase vector mesons 
should be excited.
An elegant way to describe the vector meson physics is the Hidden Local Symmetry
(HLS)~\cite{BKUY85,BKY88,HY03a} in which the vector mesons emerge as the gauge 
bosons of the HLS.
Furthermore, as the energy scale goes up, infinite number of local symmetries 
appear and the corresponding gauge fields are identified with the infinite 
vector and axial-vector mesons.
These infinite number of hidden gauge vector fields together with the pion 
field in 4-dimension (4D) can be dimensionally de-constructed to 5-dimensional 
(5D) Yang-Mills (YM) action in curved space~\cite{SS03} with the 5th dimension 
being the energy scale.

A similar situation exits in the gravity sector (that is referred to as 
``bulk" sector) of gravity/gauge (holographic) duality that comes from string 
theory.
(See, for example, a recent review addressed to hadron physicists in 
Ref.~\cite{KST12} and references therein.)
When Kaluza-Klein(KK)-decomposed to 4D and all the KK modes except the lowest 
lying vector mesons and pseudoscalar mesons are integrated out in a way 
consistent with hidden local symmetry from the bulk sector, this model is dual 
to the HLS up to $O(p^4)$~\cite{HMY06,HMY10}.
This dual (bulk-sector) model is justified in the large number of colors $N_c$, 
large 't Hooft coupling $\lambda \equiv g_{\rm YM}^2 N_c$ limit and the chiral 
limit where the quark masses vanish.
Furthermore, baryons can be described as solitons in the holographic
QCD~\cite{HRYY07,HRYY07a,HRYY07b,NSK07,HSSY07}.
The hQCD model includes only two parameters, $\lambda$ and the KK mass $M_{KK}$,
which can be fixed from meson physics.
This, therefore, enables a parameter-free calculation of the Skyrmion 
properties with 
vector mesons and provides a way to perform a systematic study on the role of 
vector mesons in the Skyrmion structure.
Here, the five-dimensional Chern-Simons (CS) term that is responsible for the 
anomalous part of the HLS Lagrangian is a topological quantity and, therefore, 
is free from the warping of the space-time.
As we shall see below, the CS term is very important to understand the role of 
the $\omega$ meson in the soliton structure.

In the literature, Skyrmion has been studied based on the $O(p^2)$ HLS 
Lagrangian as in Refs~\cite{FIKO85,IJKO85,MKW87,JJMPS88}.
This model has three parameter, $f_\pi$, $g$, and $a$, where $f_\pi$ is the 
pion decay constant, $g$ is the HLS gauge coupling constant, and $a$ is a free 
parameter in the HLS.
(See Section~\ref{sec:skyrfull} for the definition of these parameters.)
The HLS parameter $a$ is normally taken to be 
$1 \lesssim a \lesssim 2$~\cite{BKUY85,BKY88,HY03a}.
In free space $a \simeq 2$ is preferred but in hadronic medium at high 
temperature and/or density, one gets $a \simeq 1$~\cite{HY03a}.
The dependence of $a$ on circumstances hinders systematic investigation on the
properties of a single Skyrmion and baryonic matter.
For example, the soliton mass reported in Ref.~\cite{IJKO85} within a 
$\rho$-meson stabilized model is
\begin{eqnarray}
M_{\rm sol} = (667 \sim 1575) \mbox{ MeV}
\end{eqnarray}
for $1 \le a \le 4$ with $m_\pi^{} = 0$, and the pion mass effect is found to 
be small in the soliton mass.
This shows that the ambiguity in the value of $a$ results in a large 
uncertainty in the soliton mass.

Furthermore, the description of baryons as Skyrmions is supported by the large 
$N_c$ limit~\cite{Witten79b}.
In the HLS, the higher order terms such as the $O(p^4)$ terms are at $O(N_c)$ 
like the $O(p^2)$ terms.%
\footnote{The loop corrections from the $O(p^2)$ Lagrangian to the $O(p^4)$
terms are $O(N_c^0)$ and therefore are sub-dominant.
We do not consider the $O(p^6)$ Lagrangian in the present work.}
As a result, in the $N_c$ counting, these higher order terms should be taken 
into account.
However, including the higher order terms inevitably calls for more 
complicated form of the Lagrangian and uncontrollably large number of low 
energy constants.
In this paper, these constants will be determined in a controllable way by 
using a master formula.

This paper is organized as follows.
In Sec.~\ref{sec:skyrfull}, we introduce the HLS Lagrangian up to $O(p^4)$ 
including all the hWZ terms.
The soliton wave functions are constructed and the collective quantization 
method is also briefly explained.
We show a general master formula to determine the parameters of the HLS 
Lagrangian induced from a class of 5D gauge models including hQCD models in 
Sec.~\ref{sec:hlshqcd}.
In Sec.~\ref{sec:numhls}, we present our results on the Skyrmion mass and size 
as well as the moment of inertia calculated in the present work.
Here, we consider two models for the parameters, which include the HLS induced 
from the Sakai-Sugimoto (SS) model and the HLS induced from the 
Bogomol'nyi-Prasad-Sommerfield (BPS) model.
The results from these two parameter sets are discussed and compared.
Section~\ref{sec:dis} contains a summary and discussion.
The complete explicit expression for the soliton mass and the equations of 
motion for the static fields are given in Appendix~\ref{app:soliton}.
The moment of inertia and the associated equations of motion for the excited 
fields are collected in Appendix~\ref{app:moment}.


\section{Skyrmions from the hidden local symmetry}
\label{sec:skyrfull}

In order to study the role of vector mesons in Skyrmions, we first briefly 
introduce the chiral effective Lagrangian with vector mesons referring for the 
details to Refs.~\cite{HY03a,Tana93}.
Here, we consider both the iso-scalar $\omega$ meson and the iso-vector $\rho$ 
meson as well as the chiral field as explicit degrees of freedom in the theory.
These vector mesons are introduced as the gauge bosons of the HLS of the 
nonlinear sigma model~\cite{BKUY85,BKY88,HY03a}.

The full symmetry group of our effective Lagrangian is $G_{\rm full} =
[SU(2)_L\times SU(2)_R]_{\rm chiral} \times [U(2)]_{\rm HLS}$ with
$[U(2)]_{\rm HLS}$ being the HLS.
In the absence of the external sources, the HLS Lagrangian can be constructed 
by making use of the two $1$-forms $\hat{\alpha}_{\parallel\mu}$ and 
$\hat{\alpha}_{\perp\mu}$ defined by
\begin{eqnarray}
\hat{\alpha}_{\parallel\mu}^{} & = & \frac{1}{2i} \left( D_\mu \xi_R^{}
\xi_R^\dagger + D_\mu \xi_L^{} \xi_L^\dagger \right), \nonumber\\
\hat{\alpha}_{\perp\mu}^{} & = & \frac{1}{2i} \left( D_\mu \xi_R^{}
\xi_R^\dagger - D_\mu \xi_L^{} \xi_L^\dagger \right),
\end{eqnarray}
with the chiral fields $\xi_L^{}$ and $\xi_R^{}$, which are written in the
unitary gauge as
\begin{eqnarray}
\xi_L^\dagger = \xi_R^{} \equiv \xi = e^{i\pi/2f_\pi}
\quad \mbox{ with }
\pi = \bm{\pi} \cdot \bm{\tau},
\end{eqnarray}
where $\bm{\tau}$ are the Pauli matrices.
The covariant derivative is defined as
\begin{equation}
D_\mu \xi_{R,L}^{} = (\partial_\mu - i V_\mu) \xi_{R,L}^{}
\end{equation}
with $V_\mu$ being the gauge boson of the HLS.
This is the way to introduce vector mesons in the HLS, where the vector meson 
field $V_\mu$ is~\cite{BKUY85,BKY88,HY03a}
\begin{equation}
V_\mu = \frac{g}{2} (\omega_\mu + \rho_\mu)
\end{equation}
with
\begin{equation}
\rho_\mu = \bm{\rho}_\mu \cdot \bm{\tau} = \left( \begin{array}{cc}
  \rho_\mu^0  & \sqrt{2}\rho_\mu^+ \\
  \sqrt{2} \rho_\mu^- &  -\rho_\mu^0  \\
\end{array} \right) .
\end{equation}

Then one can construct the chiral Lagrangian up to $O(p^4)$ as
\begin{eqnarray}
\mathcal{L}_{\rm HLS} & = & \mathcal{L}_{\rm (2)} +
\mathcal{L}_{\rm (4)} + \mathcal{L}_{\rm anom} ,
\label{eq:Lag_HLS}
\end{eqnarray}
which is our working Lagrangian. Here,
\begin{widetext}
\begin{eqnarray}
\mathcal{L}_{\rm (2)} =
f_\pi^2 \,\mbox{Tr}\, \left( \hat{a}_{\perp\mu}
\hat{a}_{\perp}^{\mu} \right)
+ a f_\pi^2 \,\mbox{Tr}\, \left(\hat{a}_{\parallel\mu}
\hat{a}_{\parallel}^{\mu} \right)
- \frac{1}{2g^2} \mbox{Tr}\, \left( V_{\mu\nu}V^{\mu\nu} \right),
\end{eqnarray}
where $f_\pi$ is the pion decay constant, $a$ is the parameter of the HLS, 
$g$ is the vector meson coupling constant, and the field-strength tensor of 
the vector meson is
\begin{eqnarray}
V_{\mu\nu} & = & \partial_\mu V_\nu-\partial_\nu V_\mu-i[V_\mu,V_\nu].
\end{eqnarray}
In the most general form of the $O(p^4)$ Lagrangian there are several terms 
that include two traces in the flavor space such as the 
$y_{10}^{}$--$y_{18}^{}$ terms listed in Ref.~\cite{HY03a}.%
\footnote{Another example of this kind is $\mbox{Tr} 
\bigl[ \hat{\alpha}_\parallel^\mu \bigr] \mbox{Tr} 
\bigl[ \hat{\alpha}_{\parallel\mu} \bigr]$ that generates the mass 
difference between the $\rho$ and $\omega$ mesons.}
These terms are suppressed by $N_c$ compared to the other terms in the 
Lagrangian and are not considered in the present work.
Then the $O(p^4)$ Lagrangian which we study in this paper is given by
\begin{equation}
\mathcal{L}_{(4)} = \mathcal{L}_{(4)y} + \mathcal{L}_{(4)z} ,
\end{equation}
where
\begin{eqnarray}
\mathcal{L}_{(4)y} &=&
y_1^{} \mbox{Tr} \Bigl[ \hat{\alpha}_{\perp\mu}^{} \hat{\alpha}_\perp^\mu
\hat{\alpha}_{\perp\nu}^{} \hat{\alpha}_\perp^\nu \Bigr]
+ y_2^{} \mbox{Tr} \Bigl[ \hat{\alpha}_{\perp\mu}^{} \hat{\alpha}_{\perp\nu}^{}
\hat{\alpha}^\mu_\perp \hat{\alpha}^\nu_\perp \Bigr]
+ y_3^{} \mbox{Tr}
\left[ \hat{\alpha}_{\parallel\mu}^{} \hat{\alpha}_\parallel^\mu
\hat{\alpha}_{\parallel\nu}^{} \hat{\alpha}_\parallel^\nu \right]
+ y_4^{} \mbox{Tr}
\left[ \hat{\alpha}_{\parallel\mu}^{} \hat{\alpha}_{\parallel\nu}^{}
\hat{\alpha}^\mu_\parallel \hat{\alpha}^\nu_\parallel \right]
\nonumber \\ && \mbox{}
+ y_5^{} \mbox{Tr}
\left[ \hat{\alpha}_{\perp\mu}^{} \hat{\alpha}_\perp^\mu
\hat{\alpha}_{\parallel\nu}^{} \hat{\alpha}_\parallel^\nu \right]
+ y_6^{} \mbox{Tr}
\left[ \hat{\alpha}_{\perp\mu}^{} \hat{\alpha}_{\perp\nu}^{}
\hat{\alpha}^\mu_\parallel \hat{\alpha}^\nu_\parallel \right]
+ y_7^{} \mbox{Tr}
\left[ \hat{\alpha}_{\perp\mu}^{} \hat{\alpha}_{\perp\nu}^{}
\hat{\alpha}^\nu_\parallel \hat{\alpha}^\mu_\parallel \right]
\nonumber \\ && \mbox{}
+ y_8^{} \left\{
\mbox{Tr} \left[ \hat{\alpha}_{\perp\mu}^{} \hat{\alpha}_\parallel^\mu
\hat{\alpha}_{\perp\nu}^{} \hat{\alpha}_\parallel^\nu \right]
+ \mbox{Tr} \left[ \hat{\alpha}_{\perp\mu}^{} \hat{\alpha}_{\parallel\nu}^{}
\hat{\alpha}_\perp^\nu \hat{\alpha}_\parallel^\mu \right] \right\}
+ y_9^{} \mbox{Tr}
\left[ \hat{\alpha}_{\perp\mu}^{} \hat{\alpha}_{\parallel\nu}^{}
\hat{\alpha}^\mu_\perp \hat{\alpha}^\nu_\parallel \right],
\\
\mathcal{L}_{(4)z} & = &
i z_4^{} \mbox{Tr}
\Bigl[ V_{\mu\nu} \hat{\alpha}_\perp^\mu \hat{\alpha}_\perp^\nu \Bigr]
+ i z_5^{} \mbox{Tr}
\left[ V_{\mu\nu} \hat{\alpha}_\parallel^\mu \hat{\alpha}_\parallel^\nu \right].
\end{eqnarray}
\end{widetext}
In the present work, we also consider the anomalous parity hWZ terms that are 
written as
\begin{eqnarray}
\mathcal{L}_{\rm anom} & = & \frac{N_c}{16\pi^2} 
\sum_{i=1}^3 c_i \mathcal{L}_i ,
\end{eqnarray}
where
\begin{subequations}
\begin{eqnarray}
\mathcal{L}_1 & = & i \, \mbox{Tr}
\bigl[ \hat{\alpha}_{\rm L}^3 \hat{\alpha}_{\rm R}^{}
 - \hat{\alpha}_{\rm R}^3 \hat{\alpha}_{\rm L}^{} \bigr], \\
\mathcal{L}_2 & = & i \, \mbox{Tr}
\bigl[ \hat{\alpha}_{\rm L}^{} \hat{\alpha}_{\rm R}^{}
\hat{\alpha}_{\rm L}^{} \hat{\alpha}_{\rm R}^{} \bigr]  ,  \\
\mathcal{L}_3 & = & \mbox{Tr}
\bigl[ F_{\rm V} \left( \hat{\alpha}_{\rm L}^{} \hat{\alpha}_{\rm R}^{}
 - \hat{\alpha}_{\rm R}^{} \hat{\alpha}_{\rm L}^{} \right) \bigr] ,
\end{eqnarray}
\end{subequations}
in the 1-form notation with
\begin{eqnarray}
\hat{\alpha}_{L}^{} &=& \hat{\alpha}_\parallel^{} - \hat{\alpha}_\perp^{}, 
\nonumber \\
\hat{\alpha}_{R}^{} &=& \hat{\alpha}_\parallel^{} + \hat{\alpha}_\perp^{}, 
\nonumber \\
F_V &=& dV - i V^2.
\end{eqnarray}

In order to study the properties of the soliton obtained from the Lagrangian 
(\ref{eq:Lag_HLS}), we take the standard parameterization for the soliton 
configuration.
For the pion field, we use the standard hedgehog configuration,
\begin{eqnarray}
\xi(\bm{r}) = \exp\left[i\bm{\tau}\cdot\hat{\bm{r}}\frac{F(r)}{2}\right] .
\label{eq:hedgehog}
\end{eqnarray}
The configuration of the vector mesons are written as~\cite{MKW87}
\begin{equation}
\omega_{\mu} = W(r)\, \delta_{0\mu}, \quad \rho_0^{} = 0, \quad
\bm{\rho} = \frac{G(r)}{gr} \left( \hat{\bm{r}} \times \bm{\tau} \right).
\label{eq:ansatzv}
\end{equation}
For the baryon number $B=1$ solution, these wave functions satisfy the 
following boundary conditions:
\begin{eqnarray}
\begin{array}{ll}
F(0) = \pi, \qquad\qquad & F(\infty) = 0 , \\
G(0) = - 2, \qquad & G(\infty) = 0 , \\
W'(0) = 0, \qquad & W(\infty) = 0.
\end{array}
\label{eq:numprofile1}
\end{eqnarray}
Given the Lagrangian and the wave functions, it is now straightforward to 
derive the soliton mass $M_{\rm sol}$.
The explicit expression for the soliton mass is given in 
Appendix~\ref{app:soliton}.
Minimizing the soliton mass then gives the coupled equations of motion for the
wave functions $F(r)$, $W(r)$, and $G(r)$.
These are also given in Appendix~\ref{app:soliton}.

The classical configuration of the soliton obtained above should be quantized to
describe physical baryons of definite spin and isospin.
Here, we follow the standard
collective quantization method~\cite{ANW83}, which transforms the chiral field 
and the vector meson field as
\begin{eqnarray}
\xi(\bm{r}) &\to& \xi(\bm{r},t) = A(t)\, \xi(\bm{r}) A^\dagger(t), \nonumber\\
V_{\mu}(\bm{r}) &\to& V_{\mu}(\bm{r},t) = A(t)\, V_{\mu}(\bm{r}) A^\dagger(t),
\label{eq:mesoncollective}
\end{eqnarray}
where $A(t)$ is a time-dependent SU(2) matrix.
We define the angular velocity $\bm{\Omega}$ of the collective coordinate 
rotation as
\begin{eqnarray}
i \bm{\tau} \cdot \bm{\Omega} & \equiv & A^\dagger(t) \partial_0 A(t).
\label{eq:angularvelocity}
\end{eqnarray}
Under the rotation (\ref{eq:mesoncollective}), the space component of the 
$\omega$ field and the time component of the $\rho$ field, i.e., $\omega^i$ 
and $\rho^0$, get excited.
The most general forms for the vector-meson excitations are
written as~\cite{MKW87}
\begin{eqnarray}
\rho^0 (\bm{r},t) &=& A(t) \frac{2}{g}\left[ \bm{\tau} \cdot \bm{\Omega} \,
\xi_1^{}(r)
+ \hat{\bm{\tau}} \cdot \hat{\bm{r}} \, \bm{\Omega} \cdot \hat{\bm{r}} \,
\xi_2^{}(r)
\right] A^\dagger(t) , \nonumber\\
\omega^i (\bm{r},t) &=& \frac{\varphi(r)}{r} \left( \bm{\Omega} \times
\hat{\bm{r}} \right)^i ,
\label{eq:VM_excited}
\end{eqnarray}
With these wave functions the moment of inertia can be calculated and its 
explicit expression is given in Appendix~\ref{app:moment}.
It is then straightforward to obtain the Euler-Lagrange equations for the wave 
functions, $\xi_1^{}(r)$, $\xi_2^{}(r)$, and $\varphi(r)$ by minimizing the 
moment of inertia, and the results are also given in Appendix~\ref{app:moment}.
The boundary conditions imposed on the excited fields are
\begin{eqnarray}
\xi_1'(0) & = & \xi_1^{} (\infty) = 0 , \nonumber\\
\xi_2'(0) & = & \xi_2^{} (\infty) = 0 ,\nonumber\\
\varphi(0) & = & \varphi(\infty) = 0, \label{eq:BCsexcitation}
\end{eqnarray}
and $\xi_1^{}(r)$ and $\xi_2^{}(r)$ at $r=0$ satisfy the constraint,
\begin{eqnarray}
2\xi_1^{} (0) + \xi_2^{} (0) & = & 2 .
\end{eqnarray}

In the adiabatic collective quantization scheme, the baryon mass is given by
\begin{equation}
M = M_{\rm sol} + \frac{i(i+1)}{2 \mathcal{I}} 
= M_{\rm sol} + \frac{j(j+1)}{2 \mathcal{I}}
\end{equation}
where $i$ and $j$ are isospin and spin of the baryon.
Then the $\Delta$-$N$ mass difference reads
\begin{equation}
\Delta_M \equiv M_\Delta - M_N = \frac{3}{2\mathcal{I}}.
\label{eq:Delta_M}
\end{equation}

The baryonic size of a baryon should be computed by the baryon number current 
of the Skyrmion.
However, in order to intuitively see the effects of the vector mesons on the 
Skyrmion size in a simple way, here we consider the winding number and energy 
root mean square radii.
The root mean square (rms) radius of the winding number current is defined by
\begin{equation}
\langle r^2 \rangle _{W}^{1/2} = \left[ \int_0^\infty d^3r r^2 B^0(r) 
\right]^{1/2},
\end{equation}
where $B^0(r)$ is the time component of the winding number current that is 
explicitly written as
\begin{eqnarray}
B^0 & = & -\frac{1}{2\pi^2 r^2} F' \sin^2 F.
\end{eqnarray}
We define the energy root mean square radius $\langle r^2 \rangle _{E}^{1/2}$ as
\begin{equation}
\langle r^2 \rangle _{E}^{1/2} = \left[ \frac{1}{M_{\rm sol}}
\int_0^\infty d^3r r^2 M_{\rm sol}(r) \right]^{1/2},
\end{equation}
where $M_{\rm sol}(r)$ is the soliton mass (energy) density given in 
Appendix~\ref{app:soliton}.


\section{Hidden Local Symmetry induced from holographic QCD}

\label{sec:hlshqcd}

\subsection{Master formula}

In this section, following Refs.~\cite{HMY06,HMY10}, we provide a general 
master formula to determine the parameters of the HLS Lagrangian by 
integrating out the infinite towers of vector and axial-vector mesons in a 
class of hQCD models expressed by the following general 5D action:
\begin{eqnarray}
S_{\rm 5} & = & S_{\rm 5}^{\rm DBI} + S_{\rm 5}^{\rm CS},
\label{eq:actionhQCD}
\end{eqnarray}
where the 5D Dirac-Born-Infeld (DBI) part $S_{\rm 5}^{\rm DBI}$ and the
Chern-Simons (CS) part $S_{\rm 5}^{\rm CS}$ are expressed as
\begin{eqnarray}
S_{\rm 5}^{\rm DBI} & = & N_c G_{\rm YM}\int d^4 x dz
\bigg\{ - \frac{1}{2} K_1(z) \mbox{Tr}
  \left[ \mathcal{F}_{\mu\nu} \mathcal{F}^{\mu\nu} \right]
\nonumber\\
&& \qquad \qquad  \mbox{}
+ K_2(z) M_{KK}^2 \mbox{Tr} \left[ \mathcal{F}_{\mu z} \mathcal{F}^{\mu z}
\right] 
\biggr\},
\label{eq:DBI} \\
S_{\rm 5}^{\rm CS} & = & \frac{N_c}{24\pi^2} \int_{M^4\times R} w_5^{} (A).
\label{eq:SSCS}
\end{eqnarray}
where the rescaled  't Hooft coupling constant is defined as 
$G_{\rm YM} \equiv \lambda/(108\pi^3)$ and the field strength of the 5D gauge 
field%
\footnote{We use the index $M = (\mu,z)$ with $\mu = 0,1,2,3$.}
$\mathcal{A}_M$ is $\mathcal{F}_{MN} = \partial_M \mathcal{A}_N - 
\partial_N  \mathcal{A}_M - i [\mathcal{A}_M,\mathcal{A}_N]$.
Here, $K_{1,2}(z)$ are the metric functions of $z$ constrained by the 
gauge/gravity duality.
The gravity enters in the $z$ dependence of the YM coupling giving rise to the 
warping of the space.
In Eq.~(\ref{eq:SSCS}), $M^4$ and $R$ stand for the four-dimensional Minkowski 
space-time and $z$-coordinate space, respectively, and $w_5^{} (A)$ is the CS 
5-form written as
\begin{eqnarray}
w_5^{}(A) & = & \mbox{Tr} \left[ \mathcal{A} \mathcal{ F}^2 + \frac{i}{2} 
\mathcal{A}^3 
\mathcal{F} - \frac{1}{10} \mathcal{A}^5\right].
\end{eqnarray}
Here, $\mathcal{F} = d \mathcal{A} + i \mathcal{A} \mathcal{A}$ is the field 
strength of the 5D gauge field $\mathcal{A} = \mathcal{A}_\mu dx^\mu + 
\mathcal{A}_z dz$.
It should be noted that the DBI part is of $O(\lambda^1)$ while the CS term is of 
$O(\lambda^0)$ with the 't Hooft coupling constant $\lambda$.

We should stress here that as noted in Ref.~\cite{HMY10}, the structure of the 
the action (\ref{eq:actionhQCD}) is shared both by the top-down Sakai-Sugimoto 
model and the bottom-up models such as in Refs.~\cite{EKSS05,DP05} as well as 
the moose models in Ref.~\cite{SS03}, with the difference appearing only in 
the warping factors. 
This allows us to write down a ``master formula" which applies to all 
holographic models and moose construction given appropriate warping factors.

Now to induce the HLS Lagrangian from the action (\ref{eq:DBI}), we use the 
mode expansion of the 5D gauge field $\mathcal{A}_M(x,z)$ and integrate out 
all the modes except the pseudoscalar and the lowest lying vector mesons, 
which reduces $\mathcal{A}_M(x,z)$ to $A_M^{\rm integ}(x,z)$.
In the $A_z(x,z) = 0$ gauge, this implies the following 
substitution~\cite{HMY06,HMY10}:%
\footnote{As emphasized in Ref.~\cite{HMY10}, the procedure of 
``integrating out'' adopted here is different from the ``naive truncation'' 
that violates the chiral invariance.}
\begin{eqnarray}
A_\mu(x,z) & \rightarrow & A_\mu^{\rm integ}(x,z) \nonumber\\
& = & \hat{\alpha}_{\mu \perp}^{} (x) \psi_0^{}(z)
+ \left[ \hat{\alpha}_{\mu \parallel}(x) + V_\mu(x) \right]
\nonumber\\ & & \mbox{}
+ \hat{\alpha}_{\mu \parallel}^{} (x) \psi_1^{} (z) ,
\label{eq:Ainteghls}
\end{eqnarray}
where $\{\psi_{n}\}$ are eigenfunctions satisfying the following eigenvalue 
equation obtained from the action (\ref{eq:DBI}),
\begin{equation}
- K_1^{-1} (z) \partial_z \left[ K_2(z) \partial_z \psi_n^{}(z) \right] =
\lambda_n \psi_n^{}(z)  ,
\label{eq:eigenpsi}
\end{equation}
with $\lambda_n $ being the $n$-th eigenvalue ($\lambda_0 = 0$).
By substituting Eq.~(\ref{eq:Ainteghls}) into the action in Eq.~(\ref{eq:DBI}),
the HLS Lagrangian up to $O(p^4)$ can be obtained.
The explicit expressions for the LECs we need are derived as~\cite{HMY06,HMY10}
\begin{eqnarray}
f_{\pi}^2 & = & N_c G_{\rm YM}^{} M_{KK}^2  \int dz K_2(z) 
\left[ \dot{\psi}_0^{}(z)
\right]^2, \nonumber\\
a f_{\pi}^2 & = & N_c G_{\rm YM}^{} M_{KK}^2  
\lambda_1^{} \langle \psi^2_1 \rangle,
\nonumber\\
\frac{1}{g^2} & = & N_c G_{\rm YM}^{} \langle \psi_1^2 \rangle ,
\nonumber\\
y_1^{} & = & -y_2^{} = -N_c G_{\rm YM}^{} \left\langle 
\left(1 + \psi_1 - \psi_0^2 \right)^2
\right\rangle ,
\nonumber\\
y_3^{} & = & -y_4^{} = -N_c G_{\rm YM}^{} \left\langle 
\psi^2_1 \left(1 + \psi_1^{} \right)^2
\right\rangle ,
\nonumber\\
y_5^{} & = & 2 y_8^{} = -y_9^{} = -2N_c G_{\rm YM}^{} 
\left\langle \psi_1^2 \psi_0^2 \right\rangle ,
\nonumber\\
y_6^{} & = & - \left( y_5^{} + y_7^{} \right) ,
\nonumber\\
y_7^{} & = & 2N_c G_{\rm YM}^{} \left\langle \psi_1^{} 
\left ( 1 + \psi_1^{} \right)
\left(1 + \psi_1^{} - \psi_0^2 \right) \right\rangle ,
\nonumber\\
z_4^{} & = & 2N_c G_{\rm YM}^{} \left\langle \psi_1^{} 
\left( 1+\psi_1^{} - \psi_0^2 \right)
\right\rangle ,
\nonumber\\
z_5^{} & = & -2N_c G_{\rm YM}^{} \left\langle \psi_1^2 
\left( 1 + \psi_1^{} \right) \right\rangle ,
\nonumber\\
c_1^{} & = &  \left\langle\!\!\left\langle
\dot{\psi}_0^{} \psi_1^{} \left( \frac{1}{2} \psi_0^2 + \frac{1}{6} \psi_1^2
- \frac{1}{2} \right) \right\rangle\!\!\right\rangle ,
\nonumber\\
c_2^{} & = & \left\langle\!\!\left\langle
\dot{\psi}_0^{} \psi_1^{} \left( -\frac{1}{2} \psi_0^2 + \frac{1}{6} \psi_1^2
+ \frac{1}{2} \psi_1^{} + \frac{1}{2} \right) \right\rangle\!\!\right\rangle,
\nonumber\\
c_3^{} & = & \left\langle\!\!\left\langle
\frac{1}{2}\dot{\psi}_0^{} \psi_1^{2} \right\rangle\!\!\right\rangle ,
\label{eq:lecshls}
\end{eqnarray}
where $\lambda_1$ is the smallest (non-zero) eigenvalue of the eigenvalue 
equation given in Eq.~(\ref{eq:eigenpsi}), and $\langle \,\rangle$ and 
$\langle\langle\,\rangle\rangle$ are defined as
\begin{eqnarray}
\langle A \rangle & \equiv & \int_{-\infty}^{\infty}  dz K_1(z) A(z),
\nonumber\\
\langle\langle A \rangle\rangle &\equiv & \int_{-\infty}^\infty dz A(z)
\end{eqnarray}
for a function $A(z)$.
Equation (\ref{eq:lecshls}) provides the master formula for the LECs
in the HLS Lagrangian induced from general hQCD models.
Namely, one just needs to plug the warping factor and the eigenfunctions of
Eq.~(\ref{eq:eigenpsi}) into Eq.~(\ref{eq:lecshls}) to obtain the values of 
the LECs.
For example, $K_1(z) = K^{-1/3}(z)$ and $K_2(z) = K(z)$ with $K(z) = 1 + z^2$
correspond to the Sakai-Sugimoto model.

In addition to the general hQCD models, we also consider the BPS model studied
in Refs.~\cite{Sutcliffe10,Sutcliffe11} which is characterized by the flat 
space-time.
In this case, instead of solving the eigenvalue equation, the 5D gauge field
is expanded in terms of the Hermite function 
$\psi_n^{}$~\cite{Sutcliffe10,Sutcliffe11},
\begin{equation}
\psi_n^{}(z) = \frac{(-1)^{n-1}}{\sqrt{(n-1)!2^{n-1}\sqrt{\pi}}} e^{-z^2/2}
\frac{d^{n-1}}{d z^{n-1}}e^{-z^2} ,
\end{equation}
where $n \ge 1$ and $\psi_1^{}$ corresponds to the wave function of the lowest 
lying vector meson.
The wave function of the Nambu-Goldstone pseudoscalar boson is expressed in 
terms of the error function $\mbox{erf}(z)$,
\begin{equation}
\psi_0^{}(z) = \mbox{erf}(z) = \frac{2}{\sqrt{\pi}} \int_0^z e^{-\xi^2} d\xi .
\end{equation}
In the following calculation, we use the Hermit function and the error 
function as wave functions of the vector mode and pseudoscalar mode, 
respectively.
Then, the LECs of the HLS Lagrangian are determined by using the above 
$\psi_0^{}(z)$ and $\psi_1^{}(z)$ into the master formulas in 
Eq.~(\ref{eq:lecshls}) with $K(z) = 1$.

\subsection{\boldmath The $a$ independence}

In the phenomenological analysis, it is well known that the HLS parameter 
$a$ plays an important role~\cite{BKUY85,BKY88,HY03a}.
With the leading Lagrangian at $O(p^2)$, the choice of $a=2$ reproduces the
Kawarabayashi-Suzuki-Riazzudin-Fayyazudin (KSRF) relation and the $\rho$ meson
dominance in the pion electromagnetic form factor.
At $O(p^4)$, it was shown that the quantum correction enhances the infrared 
value of $a$, and, therefore, a good description of low-energy phenomenology 
can be achieved with the bare value of $a$ being $\le 2$~\cite{HY03a}.
In the holographic approach, on the other hand, the parameter $a$ is 
attributed to the normalization of the 5D wave function $\psi_1^{}(z)$, which 
cannot be determined from the homogeneous eigenvalue equation 
(\ref{eq:eigenpsi}).
As a result, it turns out~\cite{HMY10} that the physical quantities are 
independent of the parameter $a$ as far as the leading order in $N_c$ is 
concerned.%
\footnote{When we include the loop corrections which can be regarded
as a part of $1/N_c$ corrections, the $a$-dependence becomes relevant
to the physical quantities through the loop corrections.}

In order to explicitly see that any physical quantities calculated with the 
HLS Lagrangian induced from hQCD models are actually independent of the 
parameter $a$, we start with Eq.~(\ref{eq:Ainteghls}).
We first note that the vector meson mass and the pion decay constant are
related by the relation, $m_\rho^2 = a g^2 f_\pi^2$, and $m_\rho^{}$ and 
$f_\pi$ are fixed by their experimental values.
Therefore, the HLS parameter $a$ and the HLS gauge coupling $g$ are connected 
through $a g^2 = m_\rho^2 / f_\pi^2$.
Therefore, the $g$-independence of physical quantities is equivalent to their 
$a$-independence.

To see the $a$-independence explicitly, we define
\begin{eqnarray}
\widetilde{\psi}_1^{} (z) = g \psi_1^{}(z),
\end{eqnarray}
so that the new function $\widetilde{\psi}_1$ is normalized as
\begin{eqnarray}
N_c G_{\rm YM}^{} \int dz K_1(z)  \left[ \widetilde{\psi}_1(z)  \right]^2 = 1.
\end{eqnarray}
In terms of the normalized wave function $\widetilde{\psi}_1(z) $
the 5D gauge field of Eq.~(\ref{eq:Ainteghls}) is written as
\begin{eqnarray}
A_\mu^{\rm integ}(x,z) & = & \hat{\alpha}_{\mu \perp}^{} (x) \psi_0^{}(z)
+ \left[ \hat{\alpha}_{\mu \parallel}(x) + V_\mu(x)\right]
\nonumber\\
&& \mbox{}
+ \widetilde{\hat{\alpha}}_{\mu \parallel}(x) \widetilde{\psi}_1(z) , 
\label{eq:A5dHLS}
\end{eqnarray}
where $\widetilde{\hat{\alpha}}_{\mu \parallel}
= (1/g)\hat{\alpha}_{\mu \parallel}^{}$.
In terms of the radial wave functions of the soliton, $F(r)$, $W(r)$, and 
$G(r)$, we have
\begin{eqnarray}
\widetilde{\hat{\alpha}}_{\parallel}^{0}
& = & \frac{1}{g} \hat{\alpha}_{\parallel}^{0} = \frac{1}{2} W(r) , \\
\widetilde{\hat{\alpha}}_{\parallel}^{i}
& = & \frac{1}{g} \hat{\alpha}_{\parallel}^{i} = \frac12
\widetilde{G}(r)\left( \hat{\bm{r}} \times \bm{\tau} \right)^{i} ,
\end{eqnarray}
where
\begin{eqnarray}
\widetilde{G}(r) & = & \frac{1}{g r} \left[ G(r) + 1 - \cos F(r) \right] .
\end{eqnarray}
From the boundary conditions in Eq.~(\ref{eq:numprofile1}), $W(r)$ and 
$\widetilde{G}(r)$ satisfy the following boundary conditions:
\begin{eqnarray}
& & W^\prime (0) = 0  , \qquad
W (\infty) = 0 , \nonumber\\
& & \widetilde{G} (0) = 0  , \qquad \quad \;
\widetilde{G} (\infty) = 0   . \label{eq:bcalphat}
\end{eqnarray}

The coupled equations of motions for $F$, $\widetilde{G}$ and $W$ can be 
obtained by substituting the expression in Eq.~(\ref{eq:A5dHLS}) into the 
action (\ref{eq:actionhQCD}) and the minimizing the resultant energy.
This implies that there are no non-trivial boundary conditions to determine 
the absolute sizes of the vector meson contributions $W(r)$ and 
$\widetilde{G}(r)$.
The normalization of the wave functions for the vector mesons $W(r)$ and 
$\widetilde{G}(r)$ are fixed from the boundary condition of the pion 
contribution $F(r)$ through the coupled equations of motion of $F(r)$, $W(r)$, 
and $\widetilde{G}(r)$, which are independent of the gauge coupling constant 
$g$ since they are expressed in terms of the normalized $\widetilde{\psi}_1^{}$.
This can be restated as follows:
The LECs of the HLS Lagrangian are determined by substituting the expression 
in Eq.~(\ref{eq:A5dHLS}) into the action (\ref{eq:actionhQCD}), and the LECs 
are expressed in terms of normalized wave functions $\widetilde{\psi}_1^{}$ 
and $\psi_0^{}$.
Then, all the expressions in Appendix~\ref{app:soliton} can be rewritten in
terms of $F(r)$, $W(r)$ and $\widetilde{G}(r)$ without the gauge coupling 
constant $g$.
As a result, the soliton mass is free from the ambiguity of the normalization 
of $\psi_1^{}$, so that it is independent of the parameter $g$ as we shall see 
in the next section.

Similar arguments also applies to the moment of inertia.
Using Eq.~(\ref{eq:VM_excited}) one has
\begin{eqnarray}
\widetilde{\hat{a}}_{\parallel\mu} & = &
A(t)(\widetilde{\hat{a}}_{\parallel0}^{~\prime},
\widetilde{\hat{a}}_{\parallel i}^{~\prime})A^{\dag}(t),
\label{eq:hedg1formsrot}
\end{eqnarray}
where
\begin{eqnarray}
\widetilde{\hat{a}}_{\parallel}^{~0\prime} & = & 
\widetilde{\hat{a}}_{\parallel}^0
+ \left[ \bm{\tau}\cdot \bm{\Omega} \, \widetilde{\xi}_1^{}(r)
+ \bm{\tau}\cdot \hat{\bm{r}} \, \bm{\Omega}\cdot \hat{\bm{r}} \, 
\widetilde{\xi}_2^{}(r) \right], 
\nonumber\\
\widetilde{\hat{a}}_{\parallel}^{~i\prime} & = &
\widetilde{\hat{a}}_{\parallel}^{i}
+ \frac{\varphi(r)}{2r}(\bm{\Omega} \times \hat{\bm{r}})^{i} ,
\end{eqnarray}
with
\begin{eqnarray}
\widetilde{\xi}_{1}^{}(r) & = & \frac{1}{g} \left[ \xi_1^{}(r) - 1 + \cos F(r)  \right], 
\nonumber\\
\widetilde{\xi}_{2}(r) & = & \frac{1}{g} \left[ \xi_2^{} (r) + 1 - \cos F(r)
\right].
\end{eqnarray}
From Eqs.~(\ref{eq:numprofile1}) and (\ref{eq:BCsexcitation}), the boundary 
conditions for $\widetilde{\xi}_{1,2}$ read
\begin{equation}
\widetilde{\xi}_{1,2}^{~\prime} (0) = 0  , \qquad
\widetilde{\xi}_{1,2} (\infty) = 0  .
\end{equation}
Again all the equations in Appendix~\ref{app:moment} can be expressed in terms 
of $\widetilde{\xi}_{1,2}$, $\varphi$, $\widetilde{G}$, $W$, and $F$.
A similar argument to that made following Eq.~(\ref{eq:bcalphat}) yields that 
the moment of inertia should also be independent of the parameter $a$.

\subsection{\boldmath The CS term and the $\omega$ meson}

Another important point which should be addressed here with the view point
of gauge/gravity duality is that the CS term is responsible for the role of the
$\omega$ meson in the Skyrmion structure.
This can be seen by decomposing the 5D gauge field $\mathcal{A}$ into the
SU(2) and U(1) components as
\begin{eqnarray}
\mathcal{A}=A_{\rm SU(2)}+\frac 12 \tilde{A}_{\rm U(1)}.
\end{eqnarray}
Substituting this in the action of Eq.~(\ref{eq:DBI}) leads to
\begin{eqnarray}
S_{\rm 5}^{\rm DBI} &=& N_c \int d^4 x dz
\nonumber\\
&& \mbox{} \times
\bigg\{
 - \frac{1}{2} K_1(z) \left[ \mbox{Tr} \left( F_{\mu\nu} F^{\mu\nu} \right)
+ \mbox{Tr} \left( \tilde{F}_{\mu\nu} \tilde{ F}^{\mu\nu} \right)  \right]
\nonumber\\
&& \qquad \mbox{}
+ K_2(z) \left[ \mbox{Tr} \left( F_{\mu z} F^{\mu z} \right)
+ \mbox{Tr} \left( \tilde{F}_{\mu z} \tilde{F}^{\mu z} \right)
\right]
\bigg \}\ ,\nonumber\\
\label{eq:s5u1}
\end{eqnarray}
where $F_{MN}$ is the field strength of the SU(2) gauge field $A_{\rm SU(2)}$
and $\tilde{F}_{MN}$ stands for that of the U(1) gauge field 
$\tilde{A}_{\rm U(1)}$.
This explicitly shows that, without the CS term, when the 5D model is 
de-constructed from the 4D model, only the kinetic and mass terms of the 
iso-scalar vector meson $\omega$ are allowed.
This conclusion can be explicitly verified by using a specific hQCD model such 
as the SS model.

As can be read from Appendex~\ref{app:soliton}, the contribution from the 
kinetic and the mass terms of the $\omega$ meson to the soliton mass is
\begin{equation}
M_{\rm sol}^{\omega} = 4\pi \int dr \left [ - \frac{1}{2}r^2
\left( ag^2 f_\pi^2 W^2 + W'^2 \right ) \right],
\end{equation}
which gives the equation of motion of $W$ as
\begin{equation}
W'' = a g^2 f_\pi^2 W - \frac{2}{r} W'
\label{eq:EoMWnoCS}
\end{equation}
in the absence of the CS term.
By making use of the partial integration with the boundary conditions given in 
Eq.~(\ref{eq:numprofile1}), $M_{\rm sol}^{\omega}$ can be calculated as
\begin{eqnarray}
M_{\rm sol}^{\omega} &=& 4\pi \int dr \left [ - \frac{1}{2}r^2
\left( ag^2 f_\pi^2 W - \frac{2}{r} W' - W'' \right ) W \right]
\nonumber\\
&=& 0,
\end{eqnarray}
because of the equation of motion of Eq.~(\ref{eq:EoMWnoCS}).
Therefore, in the absence of the CS term, the $\omega$ field decouples from 
the other fields and $M_{\rm sol}^{\omega}$ vanishes.
This is consistent with the earlier studies on the Skyrmions stabilized by 
vector mesons~\cite{MKW87}.
As can be seen from the equation of motion of $W(r)$ given in 
Appendex~\ref{app:soliton}, the hWZ terms provide the source terms of the 
$\omega$ meson field.
Therefore, in the absence of the hWZ terms, the $\omega$ field decouples and 
does not contribute to the soliton formation.

\subsection{\boldmath The effective Skyrme parameter $e$}

Finally we estimate the Skyrme parameter $e$ in the original Skyrme
model by integrating out the isovector $\rho$ meson from the HLS.
The original Skymre model Lagrangian reads
\begin{equation}
\mathcal{L}_{\rm Sk} =
\frac{f_\pi^2}{4} \mbox{Tr} \left( \partial_\mu U \partial^\mu U^\dagger \right)
+ \frac{1}{32e^2} \mbox{Tr} \left[ \partial_\mu U U^\dagger,
\partial_\nu U U^\dagger \right]^2,
\end{equation}
where the chiral field $U$ is $U = \xi^2$ and the first term is the nonlinear 
sigma model Lagrangian that can be written as $f_\pi^2 \mbox{Tr}
(\alpha_{\perp \mu}^{} \alpha_\perp^\mu)$, where $\alpha_{\perp \mu}^{}$ is 
defined as $\hat{\alpha}_{\perp \mu}^{}$ without the vector field.
In the earlier analyses~\cite{Mei88} with the HLS Lagrangian up to $O(p^2)$, 
it is known that the Skyrme term can be obtained from the $\rho$ meson kinetic 
energy term in the limit of infinite $\rho$ meson mass. 
In this case, the Skyrme parameter $e$ becomes the $\rho$ meson coupling, so 
that we have $e = g \simeq 6$, which is close to the empirical value $e=5.45$ 
that is determined from the $\Delta$-$N$ mass difference.
In the HLS Lagrangian up to $O(p^4)$, however, we have additional contributions
from the pure $O(p^4)$ terms that lead to the Skyrme term.
Explicitly, after integrating out the $\rho$ meson, the effective Lagrangian is
obtained as
\begin{eqnarray}
\mathcal{L}_{\rm ChPT} &=&
f_\pi^2 \, \mbox{Tr} \left[ \alpha_{\perp\mu}^{} \alpha_{\perp}^{\mu} \right]
\nonumber\\  && \mbox{}
+ \left( \frac{1}{2g^2} - \frac{z_4^{}}{2}  - \frac{y_1^{} -  y_2^{}}{4} \right)
\mbox{Tr} \left[ \alpha_{\perp \mu}^{}, \alpha_{\perp \nu}^{} \right]^2
\nonumber\\ && \mbox{}
+ \frac{y_1^{} + y_2^{}}{4}\,
\mbox{Tr} \left\{\alpha_{\perp \mu}^{}, \alpha_{\perp \nu}^{} \right\}^2 ,
\label{eq:ChptfromHLS}
\end{eqnarray}
where $[\ ,\ ]$ is the commutator and $\{\ , \ \}$ is the anticommutator.
The second terms is the Skyrme term and we can read the effective Skyrme 
parameter $e$ as
\begin{eqnarray}
\frac{1}{2 e^2} & = & \frac{1}{2g^2} - \frac{z_4^{}}{2}
  - \frac{y_1^{} -  y_2^{}}{4}.
\label{eq:skyhls}
\end{eqnarray}
Since the gauge/gravity duality implies that $y_1= -y_2$, the last term of
Eq.~(\ref{eq:ChptfromHLS}) vanishes.
Using Eq.~(\ref{eq:skyhls}) and the analytic expressions for the LECs given in 
Eq.~(\ref{eq:lecshls}), the Skyrme parameter is written as
\begin{eqnarray}
\frac{1}{2 e^2} & = & \frac{N_c G_{\rm YM}}{2} 
\left\langle (1 -\psi_0^2 )^2 \right\rangle, 
\label{eq:Skyrparahls}
\end{eqnarray}
With the experimental values of the two inputs $f_\pi$ and $m_\rho$, we obtain 
the Skyrme parameter $e$ as
\begin{equation}
e \simeq 7.31.
\label{eq:SK-SS}
\end{equation}
in the SS model, while in the flat space-time case, i.e., for the BPS soliton 
model, we obtain
\begin{eqnarray}
e \simeq 10.02.
\label{eq:SkyrparaBPS}
\end{eqnarray}
These values are larger than the empirical value of the Skyrme parameter 
$e = 5.45$ because of the contributions from the $y_1^{}$, $y_2^{}$, and 
$z_4^{}$ terms that are of $O(p^4)$.

Since the moment of inertia $\mathcal{I}$ is proportional to $1/e^3$ in the 
Skyrme model, with a larger value of $e$, we have a smaller moment of inertia, 
which results in a larger mass splitting between the $\Delta$ and the nucleon 
as is verified numerically in the next Section.


\section{Numerical results for the Skyrmion}
\label{sec:numhls}

In this section, we present the results of numerical calculations on the 
Skyrmion properties in the framework of the HLS discussed in the previous 
section.
The HLS Lagrangian up to $O(p^4)$ in Eq.~(\ref{eq:Lag_HLS}) which is considered
in the present calculation contains 17 parameters, namely, $f_\pi$, $a$, $g$, 
$y_i^{}$ $(i=1,\dots 9)$, $z_4^{}$, $z_5^{}$, and $c_{1,2,3}^{}$.
We determine all these LECs through hQCD models, which are characterized by the
warping factor $K(z)$ and the wave functions $\psi_0^{}$ and $\psi_1^{}$.
Then all the LECs are obtained through the master formulas given in
Eq.~(\ref{eq:lecshls}), which contain the mass scale $M_{KK}$, the 't Hooft 
coupling $\lambda$ (or $G_{\rm YM}$), and the integrals of the warping factor 
$K(z)$ and the wave functions $\psi_0^{}$ and $\psi_1^{}$.
In the present work, we consider two hQCD models, the SS model and the BPS
model.

In hQCD models, $M_{KK}$ and $G_{\rm YM}$ are free parameters.
In the present work, we fix them by using the empirical values of $f_\pi$ and 
$m_\rho$:
\begin{eqnarray}
m_\rho & = & 775.49 \mbox{ MeV}, \nonumber\\
f_\pi & = & 92.4 \mbox{ MeV}.
\label{eq:inputs}
\end{eqnarray}
Then we have complete information to calculate all LECs of the HLS Lagrangian 
through the master formulas. 
Note, however, that the master formulas can determine only the product of 
$a g^2 = m_\rho^2/f_\pi^2$, and, therefore, $a$ or $g$ remains unfixed. 
However, as discussed in the previous section, the physical quantities are 
independent of $a$ (or $g$).
To be specific, we will first work with $a=2$, which is widely used in the 
model of the $O(p^2)$ HLS Lagrangian~\cite{BKY88,HY03a}, and then examine how 
each components of the soliton mass and the moment of inertia behave as the 
value of the HLS parameter $a$ is varied. 
This verifies numerically how the $a$ independence comes about.

In the present work, we consider three versions of the HLS model induced from 
each hQCD model.
The first version is the model that includes the pion, $\rho$ meson, and 
$\omega$ meson.
The second one is the model without the hWZ terms, i.e., the model that 
includes the pion and the $\rho$ meson.
The third one is obtained by integrating out the $\rho$ meson in the second 
version of the model. 
Therefore, this corresponds to the original Skyrme model but with the Skyrme 
parameter determined by the hQCD model. 
In this section, we will examine the three versions of the SS model and of the 
BPS model.
The obtained results will be compared with those of the $O(p^2)$ models, such 
as the $\rho$-stabilized model of Ref.~\cite{IJKO85} and ``the minimal model''
of Ref.~\cite{MKW87} that includes the $\omega$ meson in a minimal way.

\subsection{Skyrmion in the HLS induced from the Sakai-Sugimoto model}

In this subsection, we first consider the Sakai-Sugimoto model~\cite{SS04a,SS05}
to determine the LECs of the HLS Lagrangian. 
This model is characterized by the following warping factor:
\begin{eqnarray}
K_1(z) &=& \left( 1+z^2 \right)^{-1/3}, \nonumber \\
K_2(z) &=& 1 + z^2.
\end{eqnarray}
Since $M_{KK}$ and $G_{\rm YM}$ are determined by $f_\pi$ and $m_\rho$, all 
LECs except $a$ or $g$ can be determined.
This will be called HLS$_1$ model~\cite{MOYHLPR12}.
As we discussed above, we take the commonly used value $a = 2$ as a typical 
example and then we will test the results by varying the value of $a$.
The values of the LECs obtained with $a=2$ are given in the first row of 
Table.~\ref{table:LECs}.

\begin{table*}[t]
\caption{\label{table:LECs} Low energy constants of the HLS Lagrangian
at $O(p^4)$ with $a=2$.}
\begin{ruledtabular}
\centering
\begin{tabular}{clllllllll}
Model
& \qquad $y_1$ &\qquad $y_3$ & \qquad $y_5$ & \qquad $y_6$ & \qquad $z_4$ &
\qquad $z_5$ & \qquad $c_1$ & \qquad $c_2$ & \qquad $c_3$ \\
\hline
SS model & $- 0.001096$ & $ - 0.002830$ & $-0.015917$ & $+0.013712$ &
$0.010795$ & $-0.007325$ & $+0.381653$ & $-0.129602$ & $0.767374$ \\
BPS model & $- 0.071910$ & $-0.153511$ & $-0.012286$ & $-0.196545$ &
$0.090338$ & $-0.130778$ & $-0.206992$ & $+3.031734$ & $1.470210$ \\
\end{tabular}
\end{ruledtabular}
\end{table*}

Equipped with the numerical values of the LECs given in Table~\ref{table:LECs},
the equations of motion for the soliton wave function and for the soliton 
excitations can be solved numerically, which allows us to calculate the 
soliton mass and the moment of inertia.
The main results of the present work are summarized in the columns of HLS$_1$ 
of Table~\ref{table:numphy}.
The results of two models with the HLS of $O(p^2)$ are also presented for 
comparison.

\begin{table*} [t]
\centering
\caption{\label{table:numphy}
Skyrmion mass and size calculated in the HLS with the SS and BPS models with 
$a=2$.
The soliton mass $M_{\rm sol}$ and the $\Delta$-N mass difference
$\Delta_M$ are in unit of MeV while $\sqrt{\langle r^2 \rangle_W^{}}$
and $\sqrt{\langle r^2 \rangle_E^{}}$ are in unit of fm.
The column of $O(p^2) + \omega_\mu^{} B^\mu$ is ``the minimal model'' of
Ref.~\cite{MKW87} and that of $O(p^2)$ corresponds to the model of
Ref.~\cite{IJKO85}. See the text for more details.}
\begin{ruledtabular}
\begin{tabular}{c|ccc|ccc|cc}
& HLS$_1(\pi,\rho,\omega)$ & HLS$_1(\pi,\rho)$  & HLS$_1(\pi)$  &
BPS$(\pi,\rho,\omega)$  &  BPS$(\pi,\rho)$  &  BPS$(\pi)$  &
$O(p^2)+\omega_\mu^{} B^\mu$~\cite{MKW87} & $O(p^2)$~\cite{IJKO85} \\ \hline
$M_{\rm sol} $ & 1184 & 834 &  922 & 1162 & 577 &  672 &  1407 &  1026 \\
$\Delta_M$ & 448 & 1707 & 1014 & 456 & 4541 &  2613 &  259 & 1131 \\ \hline
$\sqrt{\langle r^2 \rangle_W^{}}$ &  0.433 & 0.247 &  0.309  & 0.415 & 0.164 &
0.225 &  0.540  &  0.278 \\
$\sqrt{\langle r^2 \rangle_E^{}}$ &  0.608 &  0.371 &  0.417  & 0.598 & 0.271 &
0.306 &  0.725 &  0.422 \\
\end{tabular}
\end{ruledtabular}
\end{table*}

\begin{figure}[t]
\centering
\includegraphics[width=210 pt]{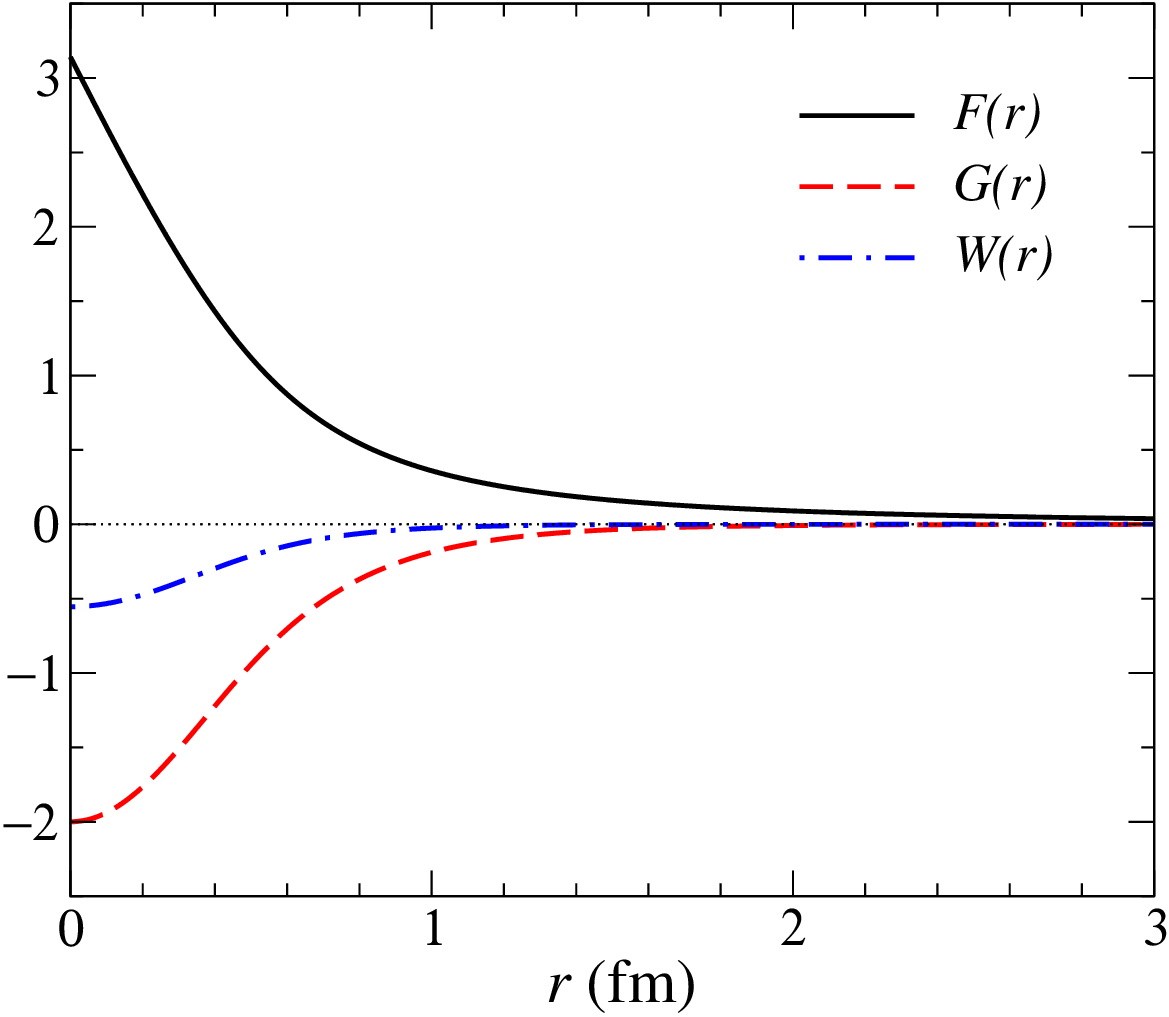}
\caption{The soliton wave functions obtained in the HLS$_1(\pi,\rho,\omega)$ 
model with $a=2$.
$F(r)$, $G(r)$, and $W(r)$ are given by the solid, dashed, and dot-dashed lines,
respectively.
Here, $F(r)$ and $G(r)$ are dimensionless, but $W(r)$ is in unit of 1/fm.}
\label{fig:profile}
\end{figure}

\begin{figure}[t]
\centering
\includegraphics[width=210 pt]{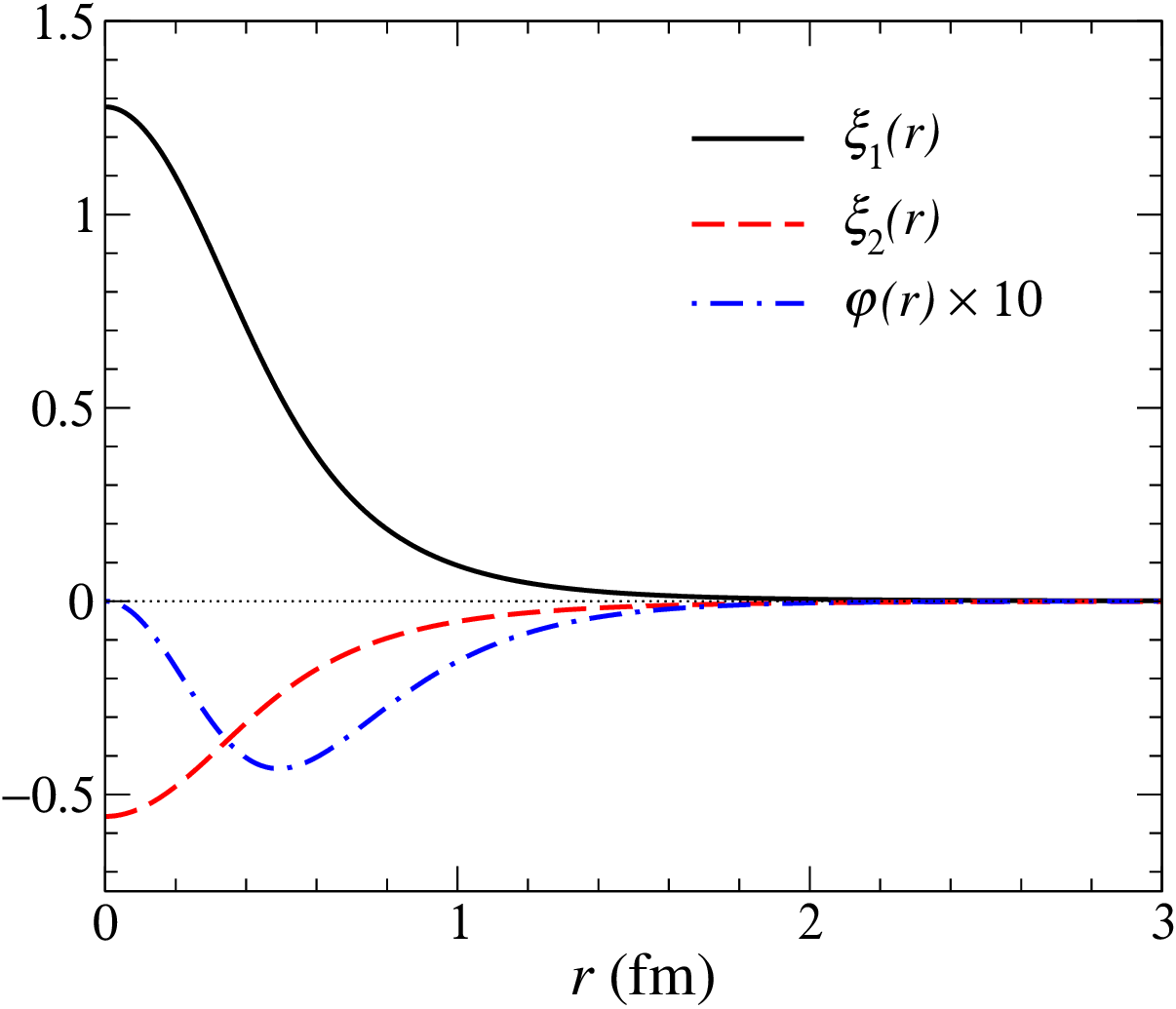}
\caption{The excitations of the soliton profile, $\xi_1^{}(r)$ (solid line), 
$\xi_2^{}(r)$ (dashed line), and $\varphi(r)$ (dot-dashed line) calculated in
the HLS$_1(\pi,\rho,\omega)$ model with $a=2$.
Here, $\xi_1^{}(r)$ and $\xi_2^{}(r)$ are dimensionless, while $\varphi(r)$ is 
in unit of fm.}
\label{fig:excitation}
\end{figure}

The obtained soliton wave functions for the HLS$_1(\pi,\rho,\omega)$ model are
shown in Fig.~\ref{fig:profile} and Fig.~\ref{fig:excitation}.
This model results in the soliton mass $M_{\rm sol} \approx 1184$~MeV and the 
moment of inertia $\mathcal{I} \approx 0.661$~fm that leads to 
$\Delta_M \approx 448$~MeV.
These numbers should be compared with the empirical values, 
$M_{\rm sol} = 867$~MeV and $\Delta_M \approx 292$~MeV.
Compared with the widely used ``minimal model" of the HLS up to
$O(p^2)$~\cite{MKW87,PRV03,LPRV03a}, this shows that the 
HLS$_1(\pi,\rho,\omega)$ model improves the soliton mass.

We then consider the HLS$_1(\pi,\rho)$ model that is constructed from the HLS 
Lagrangian without the hWZ terms.
In other words, we set $c_1^{} = c_2^{} = c_3^{} = 0$ and remove the $\omega$ 
meson mass term and its kinetic energy term in the HLS$_1(\pi,\rho,\omega)$ 
model to obtain the HLS$_1(\pi,\rho)$ model.
Therefore, this model is very similar to the model studied in 
Refs.~\cite{NSK07,NHS09}.
In addition to the soliton mass, however, we also calculate the moment of 
inertia which was not given in Refs.~\cite{NSK07,NHS09}.
And then we finally consider the HLS$_1(\pi)$ model that is defined with the 
Lagrangian (\ref{eq:ChptfromHLS}) with the Skyrme parameter given in 
Eq.~(\ref{eq:SK-SS}).
All the results are summarized in Table~\ref{table:numphy} and here are 
several comments made in order.

\begin{figure}
\centering
\includegraphics[width=230 pt]{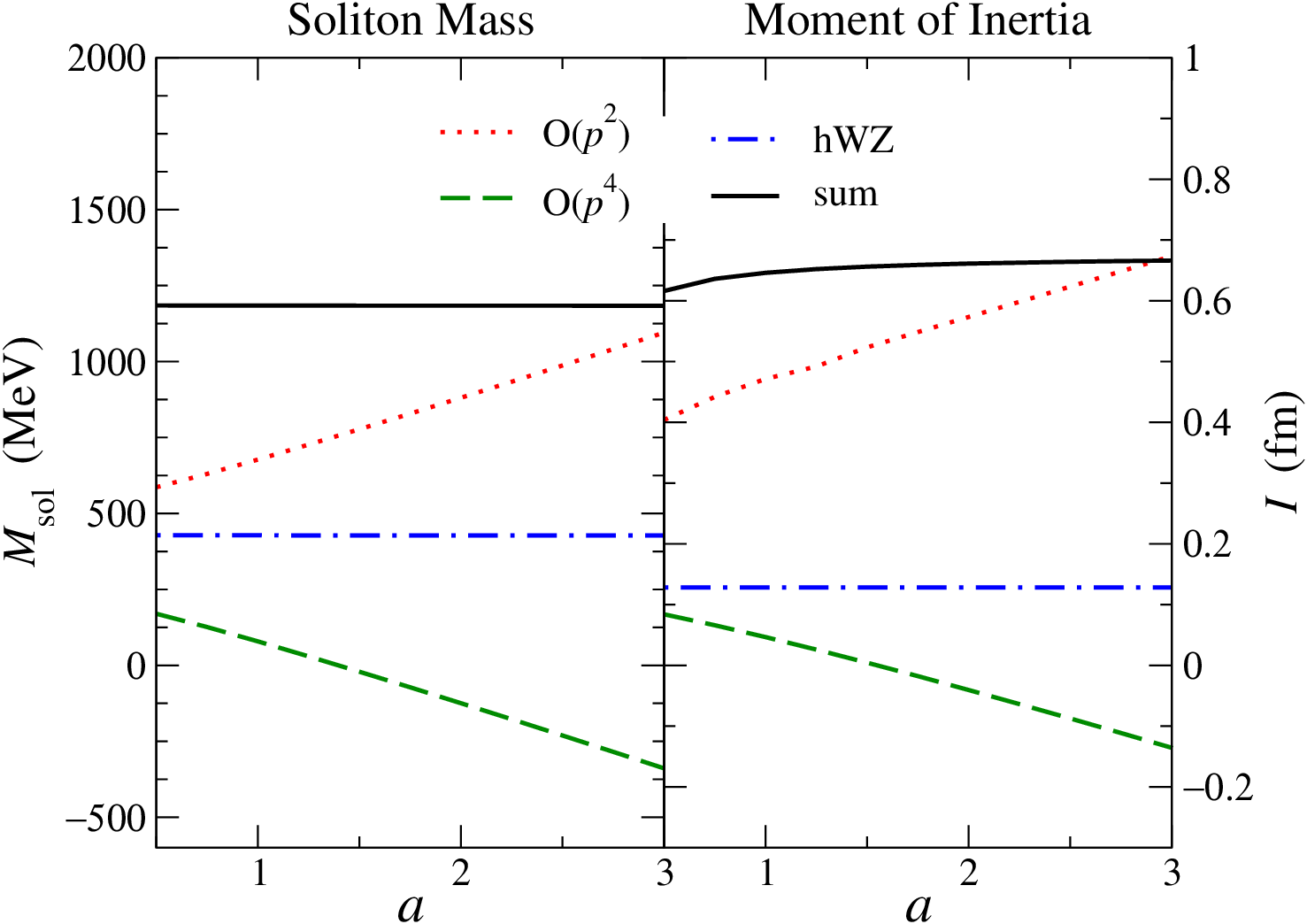}
\caption{Dependence of the soliton mass and the moment of inertia on the HLS 
parameter $a$ in the HLS$_1(\pi,\rho,\omega)$ model.
Dotted, dashed, and dot-dashed lines are the contributions from the $O(p^2)$, 
$O(p^4)$, and hWZ terms. Solid lines are their sums.}
\label{fig:HLS-a}
\end{figure}

\begin{enumerate}

\item  As claimed in the literature~\cite{NSK07,Sutcliffe10,Sutcliffe11,NHS09},
we found that the inclusion of the $\rho$ meson reduces the soliton mass.
In the present work, the soliton mass reduces from 922~MeV in the HLS$_1(\pi)$ 
to 834~MeV in the HLS$_1(\pi,\rho)$, which confirms the claim that the 
inclusion of the $\rho$ meson makes the Skyrmion closer to the BPS soliton.
However, when we include the $\omega$ meson, the soliton mass increases to 
1184~MeV.
This is in contrast to the naive expectation that including more vector mesons 
would decrease the soliton mass.
Since the $\omega$ meson interacts with the other mesons through the hWZ terms,
this observation shows the importance of the hWZ terms in the Skyrmion 
phenomenology.
The role of the $\omega$ meson in the Skyrmion mass and size can also be 
verified by comparing the soliton wave functions shown in 
Fig.~\ref{fig:HLScomp}.
This figure shows that the $\rho$ meson shrinks the soliton wave functions,
which can be seen by comparing the results from the HLS$_1(\pi)$ and the 
HLS$_1(\pi,\rho)$ models.
However, as can be seen by the dotted lines, inclusion of the $\omega$ meson 
expands the wave functions. All these behaviors can be found in the rms sizes 
$\sqrt{\langle r^2 \rangle_W^{}}$ and $\sqrt{\langle r^2 \rangle_E^{}}$ in 
Table~\ref{table:numphy}.
Therefore, we conclude that the $\rho$ meson decreases the soliton mass while 
the $\omega$ meson increases it.

\begin{figure}
\centering
\includegraphics[width=210 pt]{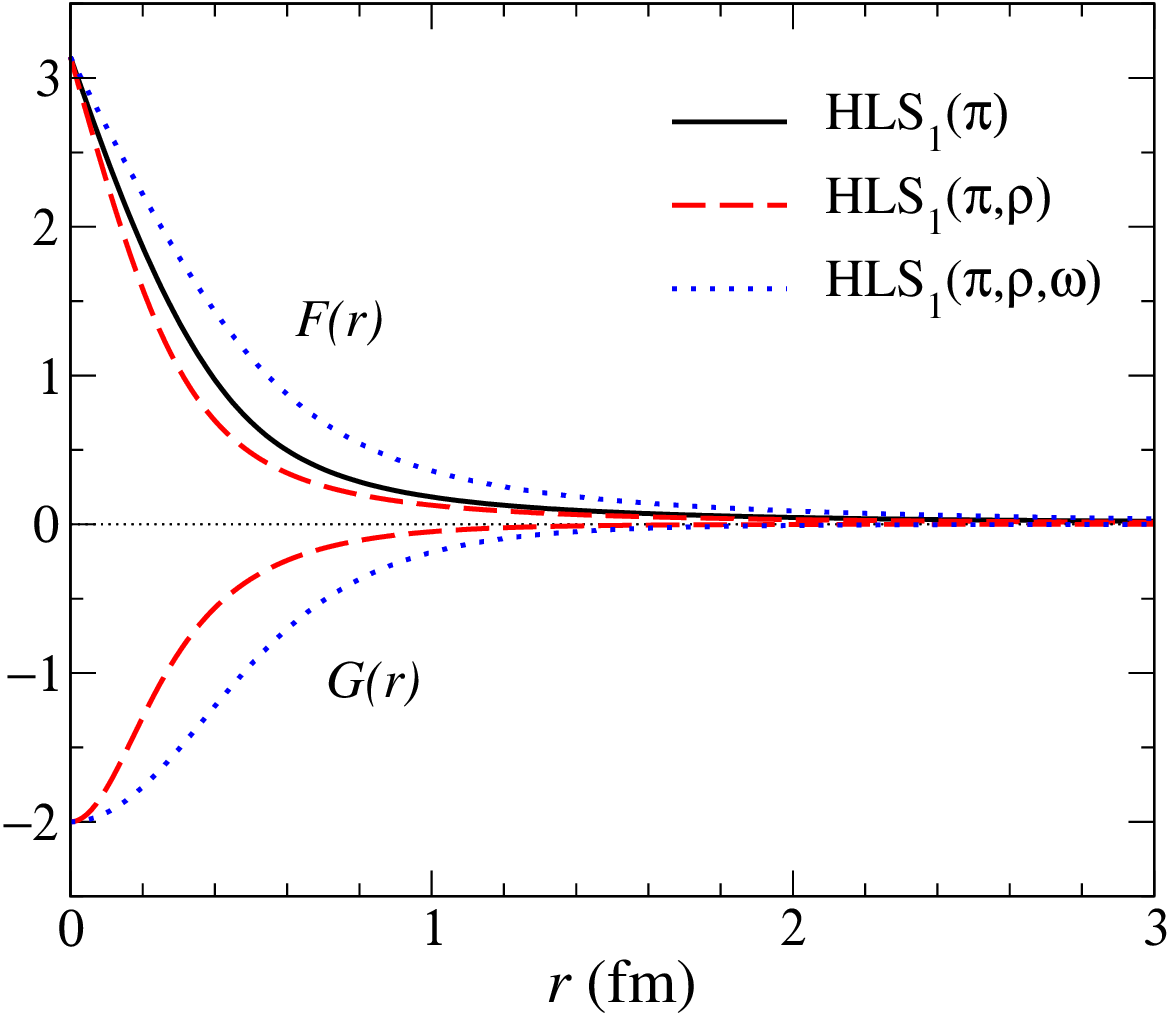}
\caption{Comparison of the soliton wave functions $F(r)$ and $G(r)$
in the three models, HLS$_1(\pi)$, HLS$_1(\pi,\rho)$, and 
HLS$_1(\pi,\rho,\omega)$, which are represented by the solid line, 
dashed lines, and dotted lines, respectively.
}
\label{fig:HLScomp}
\end{figure}

\item In the moment of inertia, or in the $\Delta$-N mass difference 
$\Delta_M$, through the collective quantization, the role of the $\rho$ and 
$\omega$ mesons are the opposite to the case of the soliton mass.
The mass difference $\Delta_M$ increases by the inclusion of the $\rho$ meson,
i.e., from 1014~MeV in the HLS$_1(\pi)$ to 1707~MeV in the HLS$_1(\pi,\rho)$,
which worsens the situation phenomenologically.
Furthermore, in the nucleon and $\Delta$ masses, the rotational energy at 
$O(1/N_c)$ is even larger than the soliton mass that is of $O(N_c)$.
This raises a serious problem to the validity of the collective quantization 
method in these models.
However, inclusion of the $\omega$ meson reduces $\Delta_M$ and the rotational 
energy appreciably.
The comparison of the soliton wave functions obtained from the $O(1/N_c)$ 
rotational energy can be found in Fig.~\ref{fig:excitationSScomp}.
This shows that $\xi_{1,2}^{}(r)$ are expanded in the HLS$_1(\pi,\rho,\omega)$ 
model than in the HLS$_1(\pi,\rho)$ model.
This leads to a larger value for the moment of inertia in the 
HLS$_1(\pi,\rho,\omega)$ model, which leads to a smaller $\Delta_M$ through 
the relation given in Eq.~(\ref{eq:Delta_M}).

\begin{figure}
\centering
\includegraphics[width=210 pt]{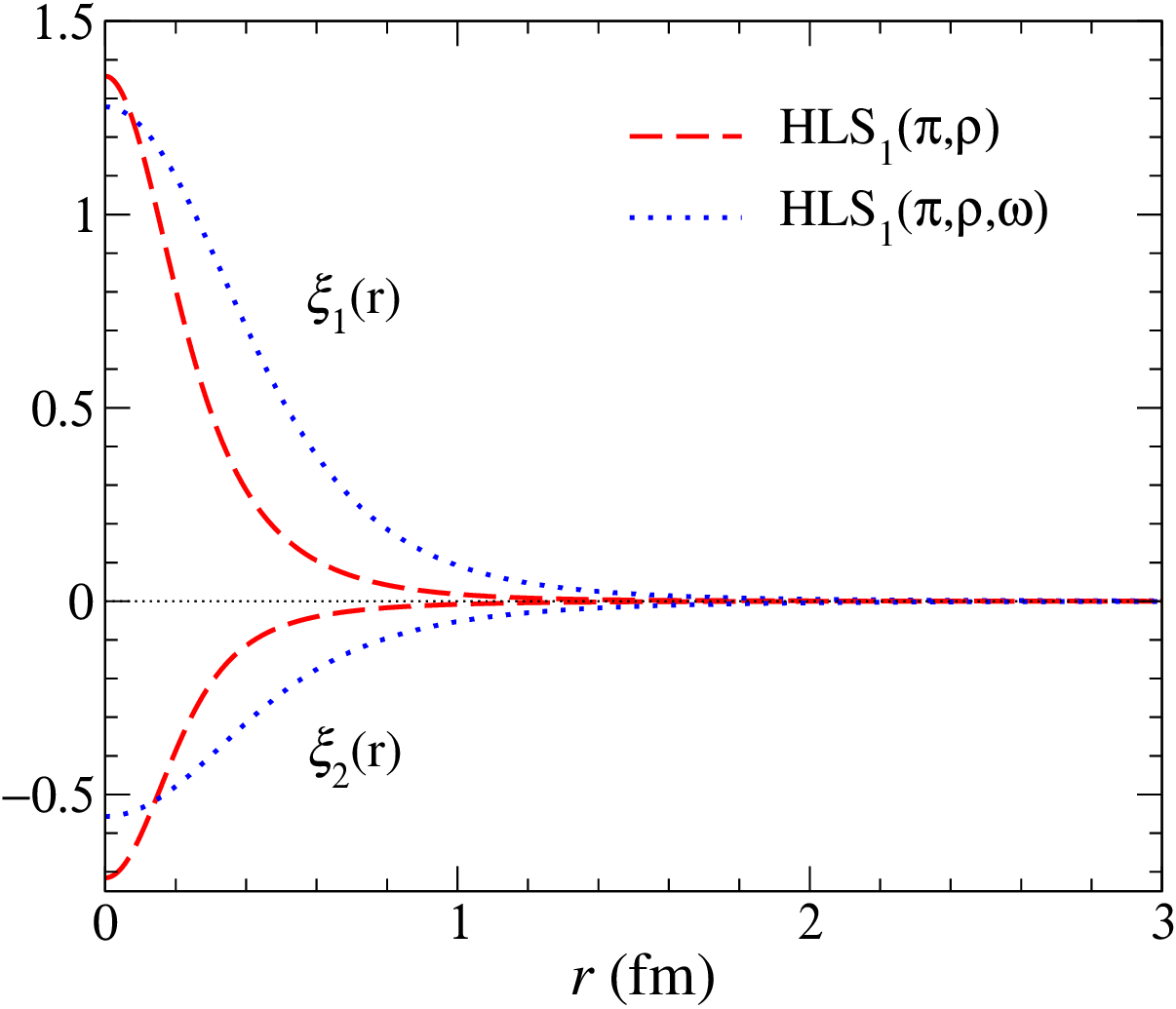}
\caption{Same as Fig.~\ref{fig:HLScomp} but for $\xi_1^{}(r)$ and 
$\xi_2^{}(r)$. }
\label{fig:excitationSScomp}
\end{figure}

\item The contributions from each term of the Lagrangian (\ref{eq:Lag_HLS}) to 
the soliton mass and the moment of inertia are also analyzed as functions of 
the HLS parameter $a$.
The results are summarized in Fig.~\ref{fig:HLS-a}.
The contributions from the $O(p^2)$ terms, $O(p^4)$ terms, and the hWZ terms 
are represented by the dotted, dashed, and dot-dashed lines, respectively, 
while the solid lines are their sums.
We first verify that the contribution from the $O(p^2)$ terms to the soliton 
mass increases with $a$.
On the contrary, the contribution from the $O(p^4)$ terms has a negative slope
with $a$ and its magnitude is smaller than the $O(p^2)$ terms,%
\footnote{The magnitude of the $O(p^4)$ contribution is about 15\% of the
$O(p^2)$ contribution at $a=2$. 
The results shown in Fig.~\ref{fig:HLS-a} show that the contribution from
the $O(p^4)$ terms is smaller than that of the $O(p^2)$ terms for $a>1$, which
shows that the power counting works for the soliton mass and the moment of inertia
at $O(N_c)$.}
which shows that this order counting is reasonable in the Skyrmion mass and 
size.
However, the contribution from the hWZ terms that are connected to the 
$\omega$ meson is highly nontrivial.
In particular, its contribution is stable as $a$ becomes smaller while the 
$O(p^2)$ contribution decreases. 
As a result, when $a \to 1$, which corresponds to the value in nuclear medium%
\footnote{$1/N_c$ corrections are expected to be highly important in medium, 
so this feature should be taken with caution.}~\cite{HY03a},
the contribution of the hWZ terms is close to that of the $O(p^2)$ terms.
This evidently shows that the role of the $\omega$ meson may be even more 
addressed in nuclear matter.
Therefore, it is highly desirable to investigate the role of the $\omega$ 
meson in more detail in Skyrmion matter.
Our analysis shows that the three components of the Skyrmion mass represented 
by the dotted, dashed, and dot-dashed lines in Fig.~\ref{fig:HLS-a} have very 
different behavior with $a$, but their sum is independent of the parameter $a$.
Similar conclusions can be drawn from the decomposition of the moment of 
inertia as well.
Here, we found a slight dependence on $a$, which might be related to the 
approximate method adopted by the collective quantization method.
More detailed studies on the quantization method is, therefore, desirable.

\end{enumerate}

All these observations show the importance of the $\omega$ meson in Skyrmions.
The $\omega$ meson increases the soliton mass and decreases the moment of 
inertia, which is exactly the opposite to the role of the $\rho$ meson.
Furthermore, only when the $\omega$ meson is included, the rotational energy 
is smaller than the soliton mass and thus the standard collective quantization 
can be justified.

\subsection{Skyrmion in the HLS induced from the BPS model}

\begin{figure}[t]
\centering
\includegraphics[width=230 pt]{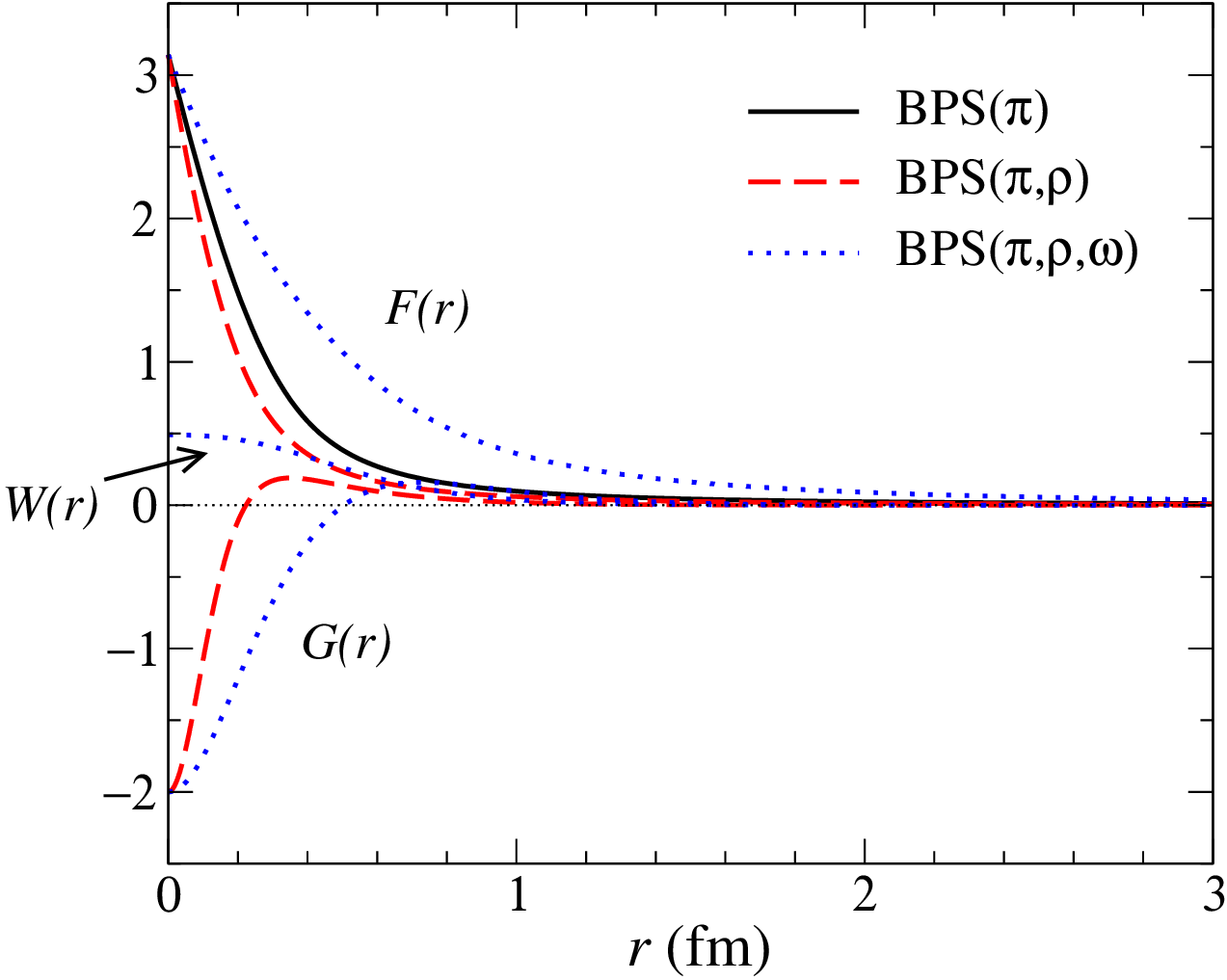}
\caption{Soliton wave functions $F(r)$ and $G(r)$ in BPS$(\pi)$,
BPS$(\pi,\rho)$, and BPS$(\pi,\rho,\omega)$ models, which are
represented by the solid line, dashed lines, and dotted lines,
respectively.
$W(r)$ is in unit of 1/fm.}
\label{fig:BPS}
\end{figure}

\begin{figure}
\centering
\includegraphics[width=210 pt]{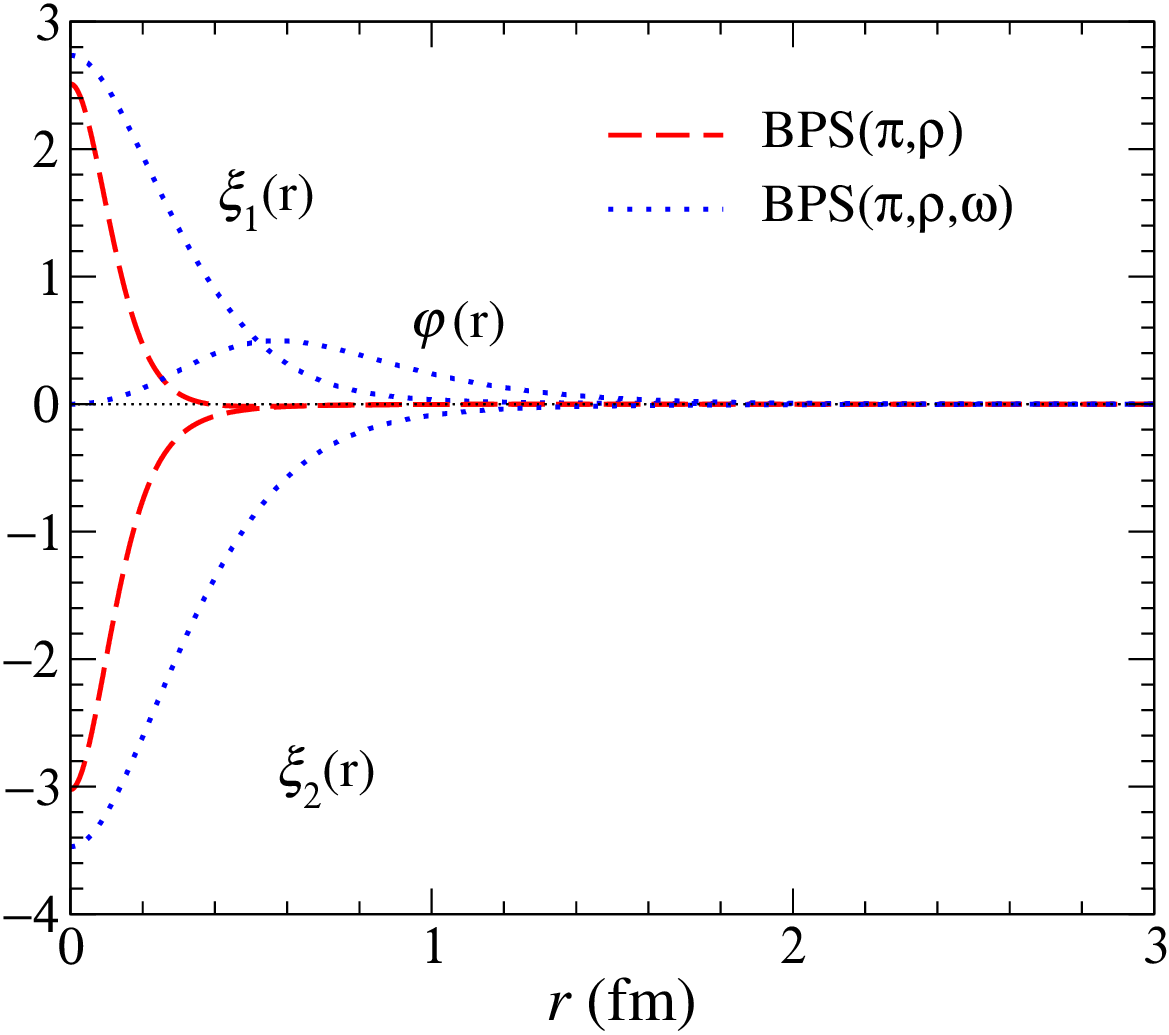}
\caption{Comparison of the soliton wave functions $F(r)$ and $G(r)$ in the 
three models, BPS$(\pi)$, BPS$(\pi,\rho)$, and BPS$(\pi,\rho,\omega)$, which 
are represented by the solid line, dashed lines, and dotted lines, respectively.
$W(r)$ is in unit of 1/fm.}
\label{fig:BPS-1}
\end{figure}

It was claimed in Refs.~\cite{Sutcliffe10,Sutcliffe11} that the BPS Skyrmion, 
i.e., the soliton in the flat space 5D YM action, has the potentially 
important feature in the Skyrmion structure.
In this subsection, we determine the LECs of the HLS with the flat space 5D 
YM action, which we call the BPS model.
To investigate the warping factor effect we consider a gauge theory in flat 
5D Minkowski space-time.
In the sense of large 't Hooft parameter $\lambda$ expansion, the flat 
space-time means that we are going to consider the $O(\lambda)$ terms since 
the warping effect is at $O(\lambda^0)$.
In flat space-time, if the infinite tower of the KK mode is included and the 
CS term is turned off, the Skyrmion solution becomes the so-called BPS Skyrmion.
In Refs.~\cite{Sutcliffe10,Sutcliffe11} the infinite tower is truncated to 
include low-lying isovector vector mesons.
Here, we include the CS term to investigate the $\omega$ meson effect.
The flat space-time is defined by
\begin{equation}
K_1 (z) = K_2 (z) = 1.
\end{equation}
As in the previous subsection, we consider the three models, namely,
BPS($\pi$), BPS($\pi,\rho$), and BPS($\pi,\rho,\omega$).
The obtained LECs and the soliton properties are also presented in 
Table~\ref{table:LECs} and \ref{table:numphy}, respectively.
The soliton wave functions in the BPS models are shown in Figs.~\ref{fig:BPS}
and \ref{fig:BPS-1} as well.

We find that the role of the vector mesons is similar to the case of the 
HLS$_1$ model.
Namely, the $\rho$ meson shrinks the soliton while the $\omega$ meson expands 
it.
Also, without the $\omega$ meson, the rotational energy, which is $O(1/N_c)$
in the baryon mass, is much larger than the soliton mass that is at $O(N_c)$.
This again raises a serious questions on the validity of the collective 
rotation in the absence of the $\omega$ meson.

\begin{figure}
\centering
\includegraphics[width=210 pt]{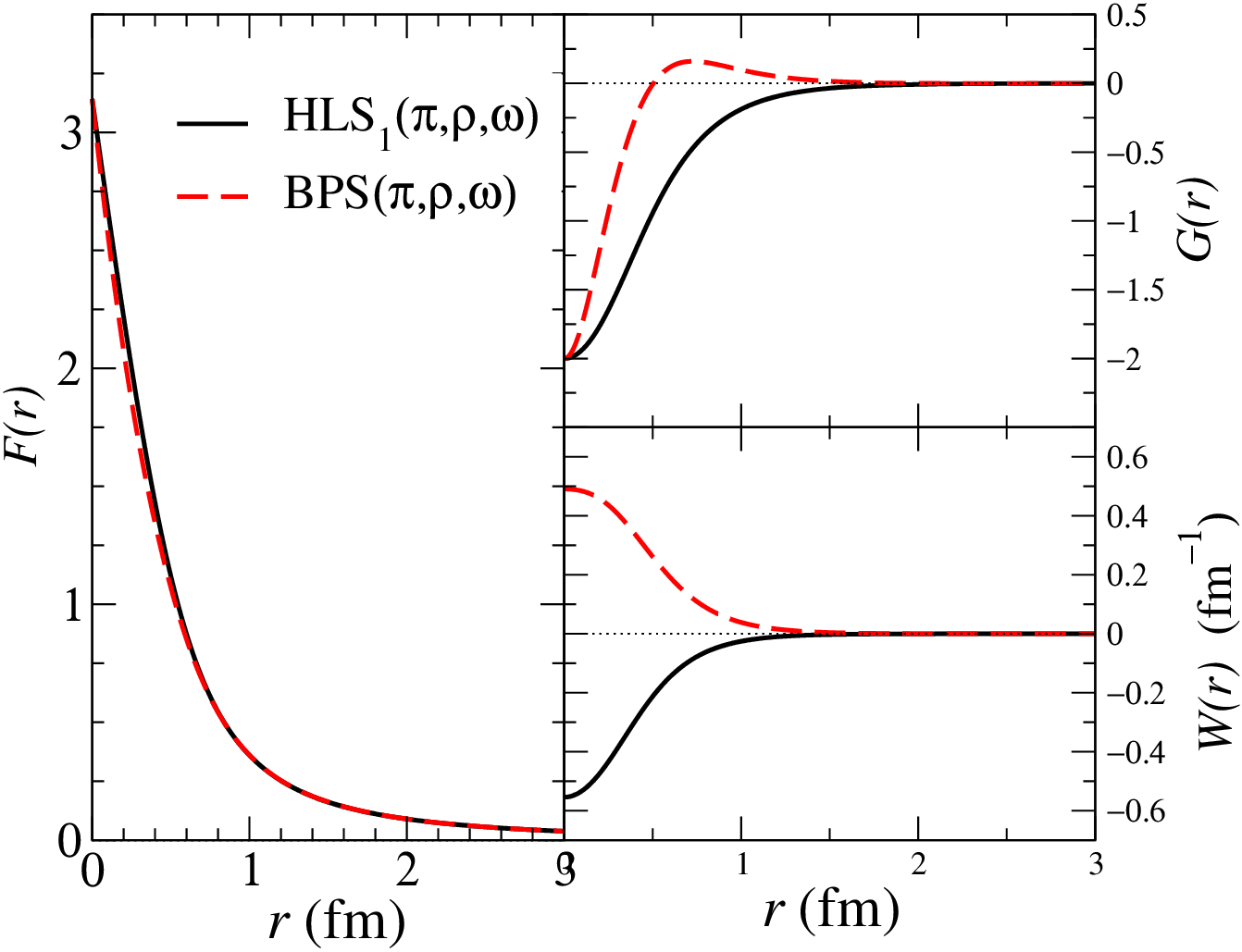}
\caption{Comparison of the sol wave functions $F(r)$, $G(r)$, and $W(r)$ 
calculated in the HLS$_1(\pi,\rho,\omega)$ and BPS($\pi,\rho,\omega$) models.}
\label{fig:SS-BPS}
\end{figure}

\begin{figure}
\centering
\includegraphics[width=210 pt]{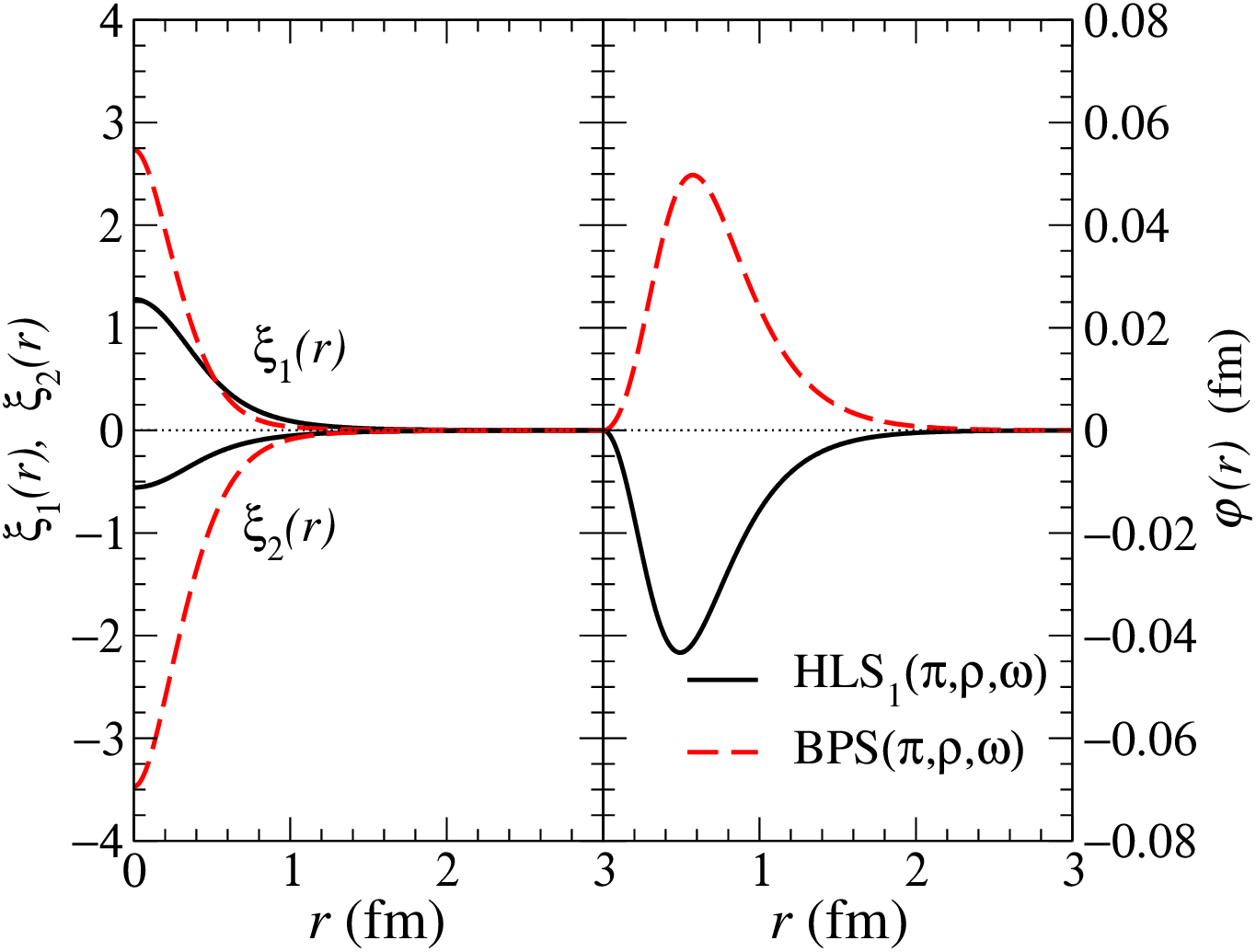}
\caption{Comparison of the sol wave functions $\xi_1^{}(r)$, $\xi_2^{}(r)$,
and $\varphi(r)$ calculated in the HLS$_1(\pi,\rho,\omega)$ and
BPS($\pi,\rho,\omega$) models.}
\label{fig:SS-BPS-1}
\end{figure}

It is also interesting to note that the soliton mass and the moment inertia 
obtained in the BPS($\pi,\rho,\omega$) model are similar to those of the 
HLS$_1(\pi,\rho,\omega)$ model, while the difference of the corresponding 
results in the models without the $\omega$ meson is quite noticeable.
Although the HLS$_1(\pi,\rho,\omega)$ and the BPS($\pi,\rho,\omega$) models 
give similar results for the soliton mass and the moment of inertia, the 
obtained soliton wave functions are very different.
To understand this coincidence, we compare the soliton wave functions in 
these two models in Figs.~\ref{fig:SS-BPS} and \ref{fig:SS-BPS-1}, which 
clearly show the difference between the two models, especially in the wave 
functions of the $\rho$ meson.
The sign difference of the wave functions of $\omega_\mu$ is due to the 
different sign in the source terms, i.e., the signs of $c_i^{}$s in 
Table~\ref{table:numphy} and it can be seen that their magnitudes are similar
in Figs.~\ref{fig:SS-BPS} and \ref{fig:SS-BPS-1}.

\begin{figure}
\centering
\includegraphics[width=230 pt]{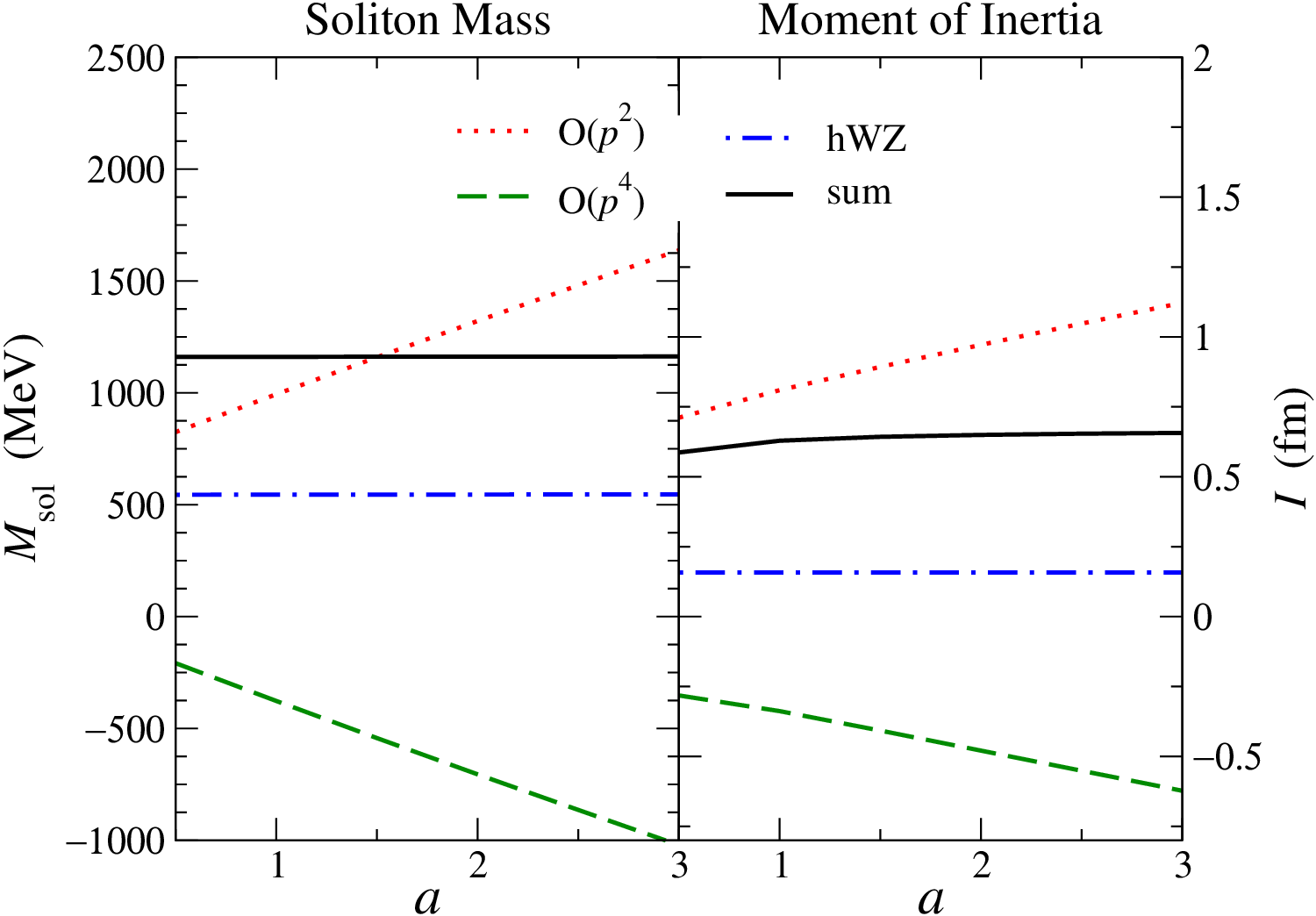}
\caption{Same as in Fig.~\ref{fig:HLS-a} but in the BPS($\pi,\rho,\omega)$
model.}
\label{fig:BPS-a}
\end{figure}

The difference between the HLS$_1(\pi,\rho,\omega)$ and the
BPS($\pi,\rho,\omega$) models can be easily seen in the breakdown of the 
soliton mass and the moment of inertia.
Shown in Fig.~\ref{fig:BPS-a} are the contribution from $O(p^2)$, $O(p^4)$, and
the hWZ terms to these physical quantities.
We first found that the dependence of these quantities on $a$ is very similar 
to that of the HLS$_1$ model.
On the contrary to the HLS$_1(\pi,\rho,\omega)$ model, however, the $O(p^4)$ 
contribution is as large as 50\% of that of the $O(p^2)$ terms at $a=2$.
Therefore, although the obtained soliton mass and the moment of inertia have 
similar values in both models, their breakdown clearly shows their difference.
Since the $O(p^4)$ contribution in the BPS model is not suppressed enough, it 
would be interesting to study the contributions from the higher order terms to 
see the convergence of the BPS model.

\section{Summary and discussion}
\label{sec:dis}

The solitonic solutions in holographic models have been
studied in the literature in terms of infinite tower of vector mesons.
In this paper, we have investigated the role of vector mesons in the Skyrmion 
properties based on the HLS Lagrangian up to $O(p^4)$ that is obtained
by integrating out the vector mesons other than the lowest $\rho$ and $\omega$
mesons in the holographic model.
In particular, by including the hWZ terms in the HLS, we have studied 
the role of the $\omega$ meson explicitly in the soliton structure.
All the LECs of the HLS Lagrangian could be determined self-consistently from 
a class of holographic QCD models by making use of the general master formulas 
(\ref{eq:lecshls}).
In the present work, we considered two hQCD models, namely, the SS model and 
the BPS model.
Equipped with the LECs of the HLS Lagrangian determined in this way, we have 
computed the Skyrmion properties and compared the results with those of the 
models of the $O(p^2)$ HLS.

The results summarized in Table~\ref{table:numphy} clearly show the important 
role of vector mesons, in particular, that of $\omega$ vector meson.
As claimed in the literature~\cite{NSK07,Sutcliffe10,Sutcliffe11,NHS09}, we 
confirmed that the inclusion of the $\rho$ meson reduces the size of the 
soliton and decreases the soliton mass so that the obtained Skyrmion mass 
becomes closer to the Bogomol'nyi bound.
It also decreases the moment of inertia, which leads to a larger value of the 
$\Delta$-N mass splitting.
However, in the model of pion only or of pion and $\rho$ meson, the obtained 
moment of inertia is very small so that the rotational energy of $O(1/N_c)$ 
becomes even larger than the soliton mass of $O(N_c)$.
This strongly raises the question on the validity of the collective rotation 
in these models.%
\footnote{See, for example, Ref.~\cite{BKS05} for the validity of the 
collective quantization in the Skyrme model.}

Then we have studied the role of the $\omega$ meson that couples to the $\pi$
and $\rho$ mesons through the hWZ terms in the HLS, which are induced from the 
CS term in hQCD models.
Contrary to the role of the $\rho$ meson, the $\omega$ meson inflates the 
soliton size and increases the soliton mass, while it reduces the rotational 
energy by increasing the moment of inertia.
Only when the $\omega$ meson is included, the rotational energy is smaller 
than the soliton mass, which validates the application of the collective 
quantization.
This shows that the $\omega$ meson has an important role in phenomenology as
well.

We further confirmed that the obtained Skyrmion properties are independent of 
the HLS parameter $a$. 
This is the consequence of the generic properties of the hQCD models related 
to the normalization of the 5D wave functions. 
Given that this relation holds in the large $N_c$ limit, the $a$-independence 
will necessarily break down in dense medium where $1/N_c$ corrections are 
expected to be important.%
\footnote{It is shown in Ref.~\cite{HY03a} that the $a$-independence
in the meson sector is broken by the loop corrections and $1 \le a \le 2$
is preferred by phenomenology. Therefore, the single soliton
properties would depend on the value of $a$ when the loop effects
are considered. This may be related to the Casimir effects and
deserves further investigation.}

It should be pointed out that, lessons from nuclear physics indicates that, 
the $\omega$ meson is responsible to the repulsive interaction which prevents 
the nuclei from collapsing and the sigma meson (not the fourth component of 
the chiral four-vector in sigma models but a scalar meson relating to the 
Casimir effects) causes attractive interaction and the near cancellation of 
the these two interactions gives the small binding energy of nuclear matter of 
$\sim 16$~MeV. 
Although it is very difficult to treat this Casimir effect given that we have 
a non-renormalizable theory, it was shown that the one-loop corrections
have nontrivial role in the properties of nucleons. 
For example, it gives the Casimir contribution of the order of $-500$~MeV, 
which goes to the right direction with a correct order of magnitude~\cite{NRZ,MW96}. 
Furthermore, in Ref.~\cite{MW96}, by adopting a chiral Lagrangian of pion, 
its effect was shown also to be important in evaluating many properties of the nucleon. 
Therefore, it would be desirable to estimate the one-loop corrections in a model
with explicit vector mesons by employing the HLS Lagrangian
employed in the present work.

Since our study shows the importance of the $\omega$ meson in the Skyrmion 
structure, it is natural to investigate its effects in the Skyrmion matter.
The approach adopted in the present paper based on the HLS with 
self-consistently determined LECs from hQCD can be extended to the study on 
dense matter.
This can be done by constructing the Skyrmions crystal lattice to determine 
the critical density at which a Skyrmion (or an instanton) transforms into two 
half-Skyrmions~\cite{LPR11,LR09a} (or half-instantons/dyons~\cite{RSZ09}).
This will be important in understanding of the equation of state for 
compact-star matter as shown in Ref.~\cite{DKLMR12}. 
As suggested in Ref.~\cite{MOYHLPR12}, a reliable treatment will require 
low-mass scalar degrees of freedom which will figure at subleading order in 
$N_c$. 
Such work is in progress and will be reported elsewhere.

\acknowledgments

We are grateful to M.~Rho for valuable discussions and critical reading of 
the manuscript.
Fruitful discussions with H.~K. Lee and B.-Y. Park are also gratefully 
acknowledged.
The work of Y.-L.M. and M.H. was supported in part by Grant-in-Aid for
Scientific Research on Innovative Areas (No.~2104) ``Quest on New
Hadrons with Variety of Flavors'' from MEXT.
Y.-L.M. was supported in part by the National Science Foundation of China (NSFC)
under Grant No.~10905060.
The work of M.H. was supported in part by the Grant-in-Aid for
Nagoya University Global COE Program
``Quest for Fundamental Principles in the Universe: From Particles
to the Solar System and the Cosmos'' from MEXT, the JSPS
Grant-in-Aid for Scientific Research (S) No.~22224003, (c)
No.~24540266.
Two of us (G.-S.Y. and Y.O.) were supported in part by the Basic Science 
Research Program through the National Research Foundation of Korea (NRF) 
funded by the Ministry of Education, Science and Technology
(Grant \mbox{No.}~2010-0009381).
The research of Y.O. was supported by Kyungpook National University Research
Fund, 2012.

\appendix

\begin{widetext}
\section{\boldmath The soliton mass and the equations of motion for 
$F(r)$, $W(r)$, and $G(r)$}
\label{app:soliton}

Using the wave functions defined in Eqs.~(\ref{eq:hedgehog}) and 
(\ref{eq:ansatzv}) and the Lagrangian in Eq.~(\ref{eq:Lag_HLS}), the soliton 
mass in the HLS up to $O(p^4)$ is obtained as
\begin{eqnarray}
M_{\rm sol} & = & 4\pi \int dr \left[ M_{(2)}(r) + M_{(4)}(r)
+ M_{\rm anom}(r) \right],
\end{eqnarray}
where $M_{(2)}$, $M_{(4)}$, and $M_{\rm anom}$ are from $\mathcal{L}_{(2)}$,
$\mathcal{L}_{(4)y} + \mathcal{L}_{(4)z}$, and $\mathcal{L}_{\rm anom}$, 
respectively.
Their explicit forms are
\begin{eqnarray}
M_{(2)}(r) &=& \frac{f_\pi^2}{2} \left( F'^2 r^2 + 2 \sin^2 F \right)
- \frac{a g^2 f_\pi^2}{2} W^2 r^2 + a f_\pi^2
\left( G + 2  \sin^2 \frac{F}{2} \right)^2
- \frac{W'^2 r^2}{2} + \frac{G'^2}{g^2} + \frac{G^2}{2g^2 r^2}
\left( G + 2 \right)^2,
\\
M_{(4)}(r) &=&
- y_1^{} \frac{r^2}{8} \left( F'^2 + \frac{2}{r^2} \sin^2 F \right)^2
- y_2^{} \frac{r^2}{8} F'^2 \left(F'^2 - \frac{4}{r^2} \sin^2 F \right)
- y_3^{} \frac{r^2}{2} \left[ \frac{g^2 W^2}{2}
- \frac{1}{r^2} \left( G + 2 \sin^2 \frac{F}{2} \right)^2 \right]^2
\nonumber \\ && \mbox{}
- y_4^{} \frac{g^2 W^2 r^2}{2} \left\{ \frac{g^2 W^2}{4}
- \frac{1}{r^2} \left( G + 2 \sin^2 \frac{F}{2} \right)^2 \right\}
+ \frac{y_5^{}}{4} \left( r^2 F'^2 + 2 \sin^2 F \right) \left[
\frac{g^2 W^2}{2}
- \frac{1}{r^2} \left( G + 2 \sin^2 \frac{F}{2} \right)^2 \right]
\nonumber \\ && \mbox{}
+ \left( y_8^{} - \frac{y_7^{}}{2} \right) \frac{\sin^2 F}{r^2}
\left( G + 2 \sin^2 \frac{F}{2} \right)^2
+ y_9^{} \left\{ \frac{g^2 W^2 r^2}{8} \left( F'^2 + \frac{2}{r^2}
\sin^2 F \right)
+ \frac{F'^2}{4} \left( G + 2 \sin^2 \frac{F}{2} \right)^2 \right\}
\nonumber \\ && \mbox{}
+ z_4^{} \left\{ G' F' \sin F + \frac{\sin^2 F}{2r^2} G(G+2) \right\}
+ \frac{ z_5^{}}{2r^2} G(G+2) \left( G + 2 \sin^2 \frac{F}{2} \right)^2,
\\
M_{\rm anom}(r) &=& \alpha_1^{} F' W \sin^2 F + \alpha_2^{} WF'
\left( G + 2 \sin^2 \frac{F}{2}
\right)^2
\nonumber \\ && \mbox{}
- \alpha_3^{} \left\{ G(G+2) WF' + 2\sin F \left[ WG' - W'
\left( G + 2 \sin^2 \frac{F}{2} \right) \right]
\right\},
\end{eqnarray}
where
\begin{eqnarray}
\alpha_1^{} = \frac{3gN_c}{16 \pi^2}  \left( c_1^{} - c_2^{} \right),
\qquad
\alpha_2^{} = \frac{gN_c}{16\pi^2} \left( c_1^{} + c_2^{} \right),
\qquad
\alpha_3^{} = \frac{gN_c}{16\pi^2} c_3^{}.
\end{eqnarray}
The Euler-Lagrange equations for $F(r)$ and $G(r)$ are obtained as
\begin{eqnarray}
\mathcal{A}_1 F'' + \mathcal{A}_2 G '' = \mathcal{B}, \nonumber \\
\mathcal{A}_3 G'' +  \mathcal{A}_4 F '' = \mathcal{D},
\end{eqnarray}
where
\begin{eqnarray}
\mathcal{A}_1 &=&
f_\pi^2 r^2
- \frac32 \left( y_1^{} + y_2^{} \right) r^2 F'^2
- \left( y_1^{} - y_2^{} \right) \sin^2 F
\nonumber \\ && \mbox{}
+ \left( y_5^{} + y_9^{} \right) \frac{g^2W^2r^2}{4}
- (y_5^{} - y_9^{}) \frac{1}{2} \left(G + 2 \sin^2 \frac{F}{2} \right)^2,
\\
\mathcal{A}_2 &=& z_4^{} \sin F,
\\
\mathcal{A}_3 &=& 1,
\\
\mathcal{A}_4 &=& \frac{g^2}{2} z_4^{} \sin F,
\end{eqnarray}
and
\begin{eqnarray}
\mathcal{B} &=& -2 f_\pi^2 r F'
+ f_\pi^2 \sin 2F
+ 2 a f_\pi^2 \sin F \left( G + 2 \sin^2 \frac{F}{2} \right)
+ f_\pi^2 m_\pi^2 r^2 \sin F
\nonumber \\ && \mbox{}
+ \left( y_1^{} + y_2^{} \right) r F'^3
+ \left( y_1^{} - y_2^{} \right) \frac{\sin 2F}{2} F'^2
- y_1^{} \frac{\sin 2F}{r^2} \sin^2 F
+ \left( y_3^{} + y_4^{} \right) g^2 W^2
\left( G + 2 \sin^2 \frac{F}{2} \right) \sin F
\nonumber \\ && \mbox{}
- y_3^{} \frac{2}{r^2} \left( G + 2 \sin^2 \frac{F}{2} \right)^3 \sin F
- \left( y_5^{} + y_9^{} \right) \frac{g^2 r W}{2} \left( W + r W' \right) F'
+ \left( y_5^{} + y_9^{} \right) \frac{g^2 W^2}{4} \sin 2F
\nonumber \\ && \mbox{}
+ \left( y_5^{} - y_9^{} \right) \left( G + 2 \sin^2 \frac{F}{2} \right)
\left( G' + \frac12 \sin F F' \right) F'
- y_5^{} \frac{\sin 2F}{2r^2} \left( G + 2 \sin^2 \frac{F}{2} \right)^2
- y_5^{} \frac{\sin^3 F}{r^2} \left( G + 2 \sin^2 \frac{F}{2} \right)
\nonumber \\ && \mbox{}
+ \left( 2 y_8^{} - y_7^{} \right) \frac{\sin 2F}{2r^2}
\left( G + 2 \sin^2 \frac{F}{2} \right)^2
+ \left( 2 y_8^{} - y_7^{} \right) \frac{\sin^3 F}{r^2}
\left( G + 2 \sin^2 \frac{F}{2} \right)
\nonumber \\ && \mbox{}
+ z_4^{} \frac{\sin 2F}{2r^2} G(G+2)
+ z_5^{} \frac{G(G+2)}{r^2} \left( G + 2 \sin^2 \frac{F}{2} \right) \sin F
\nonumber \\ && \mbox{}
- \alpha_1^{} \sin^2F W'
- \alpha_2^{} W' \left( G + 2 \sin^2 \frac{F}{2} \right)^2
- 2 \alpha_2 WG' \left( G + 2 \sin^2 \frac{F}{2} \right)
\nonumber \\ && \mbox{}
+ \alpha_3^{} \left[ G(G+2)W' + 2 \left( G + 2 \sin^2 \frac{F}{2} \right) G' W
+ 2 \cos F \left( G + 2 \sin^2\frac{F}{2} \right) W' + 2 \sin^2 F W'  \right],
\\
\mathcal{D} &=& a g^2 f_\pi^2 \left( G + 2 \sin^2 \frac{F}{2} \right)
+ \frac{1}{r^2} G(G+1)(G+2)
+ y_3^{} g^2 \left[ \frac{g^2 W^2}{2} - \frac{1}{r^2}
\left( G + 2 \sin^2 \frac{F}{2} \right)^2 \right]
 \left( G + 2 \sin^2 \frac{F}{2} \right)
\nonumber \\ && \mbox{}
+ y_4^{} \frac{g^4W^2}{2} \left( G + 2 \sin^2 \frac{F}{2} \right)
- y_5^{} \frac{g^2}{4} \left( F'^2 + \frac{2}{r^2} \sin^2 F \right)
\left( G + 2 \sin^2 \frac{F}{2} \right)
+ \left( 2 y_8^{} - y_7^{} \right) \frac{g^2}{2r^2} \sin^2 F
\left( G + 2 \sin^2 \frac{F}{2} \right)
\nonumber \\ && \mbox{}
+ y_9 \frac{g^2}{4} F'^2 \left( G + 2 \sin^2 \frac{F}{2} \right)
- z_4^{} \frac{g^2}{2} \cos F F'^2 + z_4^{} \frac{g^2}{2r^2} \sin^2 F (G+1)
\nonumber \\ && \mbox{}
+ z_5^{} \frac{g^2}{2r^2} \left[ (G+1)\left( G + 2 \sin^2 \frac{F}{2} \right)
+ G(G+2) \right]
\left( G + 2 \sin^2 \frac{F}{2} \right)
\nonumber \\ && \mbox{}
+ \alpha_2^{} g^2 WF' \left( G + 2 \sin^2 \frac{F}{2} \right)
+ \alpha_3 g^2 \left[ 2W' \sin F - WF' \left( G + 2 \sin^2 \frac{F}{2} \right)
\right].
\end{eqnarray}

The equation of motion of $W$ reads
\begin{eqnarray}
W'' &=& - \frac{2}{r} W' + a g^2 f_\pi^2 W
+ \left( y_3^{} + y_4^{} \right) g^2 W
\left[ \frac{g^2 W^2}{2} - \frac{1}{r^2}
\left( G + 2 \sin^2 \frac{F}{2} \right)^2 \right]
\nonumber \\ && \mbox{}
- \left( y_5^{} + y_9^{} \right) \frac{g^2 W}{4}
\left( F'^2 + \frac{2}{r^2} \sin^2F \right)
- \alpha_1^{} \frac{\sin^2 F}{r^2} F'
- \alpha_2^{} \frac{F'}{r^2} \left( G + 2 \sin^2 \frac{F}{2} \right)^2
\nonumber \\ && \mbox{}
+ \frac{\alpha_3^{}}{r^2} \left[ \left( G^2 + 2G + 2 \sin^2 F \right) F'
+ 2 \cos F \left( G + 2 \sin^2 \frac{F}{2} \right) F' + 4 \sin F G' \right].
\end{eqnarray}
This evidently shows that the hWZ terms, i.e, the $c_i^{}$ terms, are the
source terms of $W(r)$.


\section{Moment of inertia and equations of motion of the excited fields }
\label{app:moment}

When the collective rotation is introduced, the Lagrangian can be written as
\begin{equation}
L = - M_{\rm sol} + \mathcal{I} \, \mbox{Tr} \left( \dot{A} \dot{A}^\dagger
\right),
\end{equation}
where the moment of inertia $\mathcal{I}$ is summarized as
\begin{eqnarray}
\mathcal{I} & = & 4\pi \int dr \left[ \mathcal{I}_{(2)}(r) +
\mathcal{I}_{(4)}(r)
+ \mathcal{I}_{\rm anom}(r) \right].
\end{eqnarray}
The contributions from $\mathcal{L}_{(2)}$, $\mathcal{L}_{(4)y} +
\mathcal{L}_{(4)z}$, and $\mathcal{L}_{\rm anom}$, are represented by
$\mathcal{I}_{(2)}(r)$, $\mathcal{I}_{(4)}(r)$, and
$\mathcal{I}_{\rm anom}(r)$, respectively.
We further write
\begin{equation}
\mathcal{I}_{(4)}= \sum_i y_i^{} \mathcal{I}_{y_i^{}} +
\sum_i z_i^{} \mathcal{I}_{z_i^{}}.
\end{equation}
The moment of inertia from $\mathcal{L}_{(2)}$ is obtained as
\begin{eqnarray}
\mathcal{I}_2 (r)  &=& \frac23 f_\pi^2  r^2\sin^2 F
+ \frac13 a f_\pi^2 r^2 \left[ \left( \xi_1^{} + \xi_2^{} \right)^2
+ 2 \left( \xi_1^{} - 2 \sin^2 \frac{F}{2} \right)^2 \right]
\nonumber \\ && \mbox{}
- \frac16 a g^2 f_\pi^2 \varphi^2
- \frac16 \left( \varphi'^2 + \frac{2\varphi^2}{r^2} \right)
+ \frac{r^2}{3g^2} \left( 3 \xi_1'^2 + 2 \xi_1' \xi_2' + \xi_2'^2 \right)
\nonumber \\ && \mbox{}
+ \frac{4}{3g^2} G^2 \left( \xi_1^{} - 1 \right)
\left( \xi_1^{} + \xi_2^{} - 1 \right)
+ \frac{2}{3g^2} \left( G^2 + 2 G+2 \right) \xi_2^2.
\end{eqnarray}
Each term of $\mathcal{I}_{(4)}$ is calculated as
\begin{eqnarray}
\mathcal{I}_{y_1^{}} (r)  &=& - \frac13 r^2 \sin^2 F
\left( F'^2 + \frac{2}{r^2} \sin^2 F \right),
\\
\mathcal{I}_{y_2^{}} (r)  &=& \frac13 r^2 \sin^2 F F'^2 ,
\\
\mathcal{I}_{y_3^{}} (r)  &=& - \frac{1}{12} g^2 \varphi^2
\left[ g^2 W^2 - \frac{4}{r^2} \left( G + 2 \sin^2 \frac{F}{2} \right)^2 \right]
+ \frac23 g^2 W \varphi \left( G + 2 \sin^2 \frac{F}{2} \right) \left( \xi_1^{} - 2 \sin^2
\frac{F}{2} \right)
\nonumber \\ && \mbox{}
+ \left[ \frac12 r^2 g^2 W^2 - \frac13 \left( G + 2 \sin^2 \frac{F}{2}
\right)^2 \right]
\left[ \left( \xi_1^{} + \xi_2^{} \right)^2 +
2 \left( \xi_1^{} - 2 \sin^2 \frac{F}{2} \right)^2 \right],
\\
\mathcal{I}_{y_4^{}} (r)  &=& \frac{r^2}{2} g^2 W^2 \left[ 
\left( \xi_1^{} + \xi_2^{} \right)^2
+ 2 \left( \xi_1^{} - 2 \sin^2 \frac{F}{2} \right)^2 \right]
- \frac{1}{12} g^2 W \varphi \left[ g^2 W \varphi -
8 \left( G + 2 \sin^2 \frac{F}{2} \right)
\left( \xi_1^{} - 2 \sin^2 \frac{F}{2} \right) \right]
\nonumber \\ && \mbox{}
+ \frac13 \left( G + 2 \sin^2 \frac{F}{2} \right)^2
\left[ \frac{g^2 \varphi^2}{r^2} +
\left( \xi_1^{} + \xi_2^{} \right)^2 \right],
\\
\mathcal{I}_{y_5^{}} (r)  &=& \frac16 \sin^2 F
\left[ r^2 g^2 W^2 - 2 \left( G + 2 \sin^2 \frac{F}{2} \right)^2 \right]
\nonumber \\ && \mbox{}
- \frac{r^2}{12} \left( F'^2 + \frac{2}{r^2} \sin^2 F \right)
\left[ 2 \left( \xi_1^{} - 2 \sin^2 \frac{F}{2} \right)^2
+ \left( \xi_1^{} + \xi_2^{} \right)^2
- \frac{g^2 \varphi^2}{2r^2} \right],
\\
\mathcal{I}_{y_6^{}} (r)  &=& \frac{1}{6} \sin^2 F
\left( rgW - \frac{g\varphi}{2r} \right)^2,
\\
\mathcal{I}_{y_7^{}} (r)  &=& \frac{1}{6} \sin^2 F
\left[ \left( rgW - \frac{g\varphi}{2r} \right)^2
+ 4 \left( G + 2 \sin^2 \frac{F}{2} \right)
\left( \xi_1^{} - 2 \sin^2 \frac{F}{2} \right) \right],
\\
\mathcal{I}_{y_8^{}} (r)  &=& \frac{1}{3} \sin^2 F
\left[ \left( rgW - \frac{g\varphi}{2r} \right)^2
- 4 \left( G + 2 \sin^2 \frac{F}{2} \right)
\left( \xi_1^{} - 2 \sin^2 \frac{F}{2} \right) \right],
\\
\mathcal{I}_{y_9^{}} (r)  &=& \frac{r^2}{6} g^2 W^2 \sin^2 F
+ \frac{r^2}{6} F'^2 \left( \xi_1^{} - 2 \sin^2 \frac{F}{2} \right)^2
\nonumber \\ && \mbox{}
- \frac{r^2}{12} \left( F'^2 - \frac{2}{r^2} \sin^2F \right)
\left( \xi_1^{} + \xi_2^{} \right)^2
+ \frac{1}{24} g^2 \varphi^2 \left( F'^2 + \frac{2}{r^2} \sin^2 F \right),
\end{eqnarray}
and
\begin{eqnarray}
\mathcal{I}_{z_4^{}} (r)  &=& \frac23 \sin^2 F
\left[ G \left( 1 - \xi_1^{} \right) + \xi_2^{} \right]
- \frac23 r^2 \sin F F' \xi_1',
\\
\mathcal{I}_{z_5^{}} (r)  &=& - \frac23 \left( G + 2 \sin^2 \frac{F}{2} \right)
\Biggl\{ \left[ G \left( 1 - \xi_1^{} - \xi_2^{} \right) - \xi_2^{} \right]
\left( \xi_1^{} + \xi_2^{} \right)
+ \left[ G \left( 1 - \xi_1^{} \right) + \xi_2^{} \right]
\left( \xi_1^{} - 2 \sin^2 \frac{F}{2} \right)
\Biggr\},
\end{eqnarray}
The hWZ terms give
\begin{eqnarray}
\mathcal{I}_{\rm anom} (r)  &=& \frac{gN_c}{8\pi^2}
\left( c_1^{} - c_2^{} \right) \varphi F' \sin^2 F
\nonumber \\ && \mbox{}
- \frac{gN_c}{24\pi^2} \left( c_1^{} + c_2^{} \right) \varphi F'
\left( G + 2 \sin^2 \frac{F}{2} \right)
\left( \xi_1^{} - 2 \sin^2 \frac{F}{2} \right)
\nonumber \\ && \mbox{}
+ \frac{gN_c}{24\pi^2} c_3^{} \Biggl\{ \varphi F'
\left( G \xi_1 - G - \xi_1 - 2 \xi_2 \right)
+ \varphi \sin F \left( \xi_1' - G' \right)
+ \varphi' \sin F \left( G - \xi_1 + 4 \sin^2 \frac{F}{2} \right) \Biggr\}.
\end{eqnarray}

Then the equations of motion for $\xi_1^{}$, $\xi_2^{}$, and $\varphi$ are
obtained as
\begin{eqnarray}
\xi_1'' &=& - \frac{2}{r} \xi_1' + a g^2 f_\pi^2 \left( \xi_1^{}
- 2 \sin^2 \frac{F}{2} \right)
+ \frac{G^2}{r^2} \left( \xi_1^{} - 1 \right) - \frac{2}{r^2} (G+1) \xi_2^{}
+ \frac{3g^2}{4r^2} \left( \mathcal{F}_1 - \mathcal{F}_2 \right)
\nonumber \\ && \mbox{}
- \frac{g^3 N_c}{32 \pi^2 r^2} \left( c_1^{} + c_2^{} \right) \varphi F'
\left( G + 2 \sin^2 \frac{F}{2}
\right)
- \frac{g^3 N_c}{32 \pi^2 r^2} c_3^{}
\left[ 2 \varphi' \sin F - \varphi F'
\left( G + 2 \sin^2 \frac{F}{2} \right) \right],
\\
\xi_2'' &=& - \frac{2}{r} \xi_2' + a g^2 f_\pi^2
\left( \xi_2^{} + 2 \sin^2 \frac{F}{2} \right)
+ \frac{G^2}{r^2} \left( \xi_1^{} + 2 \xi_2^{} - 1 \right)
+ \frac{6}{r^2} (G+1) \xi_2^{}
+ \frac{3g^2}{4r^2} \left( 3 \mathcal{F}_2 - \mathcal{F}_1 \right)
\nonumber \\ && \mbox{}
+ \frac{g^3 N_c}{32 \pi^2 r^2} \left( c_1^{} + c_2^{} \right)
\varphi F' \left( G + 2 \sin^2 \frac{F}{2}
\right)
+ \frac{g^3 N_c}{32 \pi^2 r^2} c_3^{} 
\left[ 2 \varphi' \sin F - \varphi F' \left( G + 5 - \cos F \right)
\right],
\\
\varphi'' &=& \frac{2}{r^2} \varphi + ag^2 f_\pi^2 \varphi - 3 \mathcal{F}_3
- \frac{3gN_c}{8\pi^2} \left( c_1^{} - c_2^{} \right) F' \sin^2 F
+ \frac{gN_c}{8\pi^2} \left( c_1^{} + c_2^{} \right) F'
\left( G + 2 \sin^2 \frac{F}{2} \right)
\left( \xi_1^{} - 2 \sin^2 \frac{F}{2} \right)
\nonumber \\ && \mbox{}
+  \frac{gN_c}{8\pi^2} c_3^{} \Biggl\{ 2 \sin F \left( G' - \xi_1' \right)
+ F' \left[ G (1+ \cos F)  - \xi_1^{} \left( G - 2 \sin^2 \frac{F}{2} \right)
+ 2 \xi_2^{} + 3 \sin^2 F - 4 \sin^4 \frac{F}{2} \right] \Biggr\},
\nonumber \\
\end{eqnarray}
where
\begin{eqnarray}
\mathcal{F}_1 &=& y_3^{} \Biggl\{ \frac23 g^2 W \varphi
\left( G + 2 \sin^2 \frac{F}{2} \right)
+ \left[ r^2 g^2 W^2 - \frac23
\left( G + 2 \sin^2 \frac{F}{2} \right)^2 \right]
\left( 3 \xi_1^{} + \xi_2^{} - 4 \sin^2 \frac{F}{2} \right) \Biggr\}
\nonumber \\ && \mbox{}
+ y_4^{} \Biggl\{ r^2 g^2 W^2 \left( 3 \xi_1^{} + \xi_2^{}
- 4 \sin^2 \frac{F}{2} \right)
+ \frac23 g^2 W \varphi \left( G + 2 \sin^2 \frac{F}{2} \right)
+ \frac23 \left( G + 2 \sin^2 \frac{F}{2} \right)^2
\left( \xi_1^{} + \xi_2^{} \right) \Biggr\}
\nonumber \\ && \mbox{}
- y_5^{} \frac{r^2}{6} \left( F'^2 + \frac{2}{r^2} \sin^2 F \right)
\left( 3 \xi_1^{} + \xi_2^{}
- 4 \sin^2 \frac{F}{2} \right)
+ \left( y_7^{} - 2 y_8^{} \right) \frac23 \sin^2 F
\left( G + 2 \sin^2 \frac{F}{2} \right)
\nonumber \\ && \mbox{}
+ y_9^{} \frac{r^2}{3} \left\{ F'^2
\left( \xi_1^{} - 2 \sin^2 \frac{F}{2} \right)
- \frac12 \left( F'^2 - \frac{2}{r^2} \sin^2 F \right)
\left( \xi_1^{} + \xi_2^{} \right) \right\}
\nonumber \\ && \mbox{}
+ z_4^{} \frac23 \left\{ -G \sin^2 F + r^2 \cos F F'^2
+ r^2 \sin F \left( F'' + \frac{2}{r} F' \right)
\right\}
\nonumber \\ && \mbox{}
- z_5^{} \frac43 G \left( G + 2 \sin^2 \frac{F}{2} \right)
\left( 1 - 2 \xi_1^{} - \xi_2^{}
+ \sin^2 \frac{F}{2} \right),
\\
\mathcal{F}_2 &=& y_3^{} \left[ r^2 g^2 W^2
- \frac23 \left( G + 2 \sin^2 \frac{F}{2} \right)^2 \right]
\left( \xi_1^{} + \xi_2^{} \right)
+ y_4^{} \left[ r^2 g^2 W^2
+ \frac23 \left( G + 2 \sin^2 \frac{F}{2} \right)^2 \right]
\left( \xi_1^{} + \xi_2^{} \right)
\nonumber \\ && \mbox{}
- y_5^{} \frac{r^2}{6} \left( F'^2 + \frac{2}{r^2}\sin^2 F \right) \left( \xi_1^{} + \xi_2^{} \right)
- y_9^{} \frac{r^2}{6} \left( F'^2 - \frac{2}{r^2}\sin^2 F \right) \left( \xi_1^{} + \xi_2^{} \right)
\nonumber \\ && \mbox{}
+ z_4^{} \frac23 \sin^2 F
- z_5^{} \frac23 \left( G + 2 \sin^2 \frac{F}{2} \right)
\left[ G \left( 1 - 2 \xi_1^{} \right)
- 2 (G+1) \xi_2^{} - 2 \sin^2 \frac{F}{2} \right],
\\
\mathcal{F}_3 &=& - y_3^{} \frac16 g^2 \varphi
\left[ g^2 W^2 - \frac{4}{r^2}
\left( G + 2 \sin^2 \frac{F}{2} \right)^2 \right]
+ y_3^{} \frac23 g^2 W \left( G + 2 \sin^2 \frac{F}{2} \right)
\left( \xi_1^{} - 2 \sin^2 \frac{F}{2} \right)
\nonumber \\ && \mbox{}
- y_4^{} \frac16 g^2 W \left[ g^2 W \varphi
- 4 \left( G + 2 \sin^2 \frac{F}{2} \right)
\left( \xi_1^{} - 2 \sin^2 \frac{F}{2} \right) \right]
+ y_4^{} \frac23 \frac{g^2 \varphi}{r^2}
\left( G + 2 \sin^2 \frac{F}{2} \right)^2
\nonumber \\ && \mbox{}
+ \left( y_5^{} + y_9^{} \right) \frac{1}{12} g^2 \varphi
\left( F'^2 + \frac{2}{r^2} \sin^2 F \right)
- \left( y_6^{} + y_7^{} + 2 y_8^{} \right) \frac{g}{6} \sin^2 F
\left( gW - \frac{g\varphi}{2r^2} \right).
\end{eqnarray}

\end{widetext}

\end{document}